\documentclass[aps,pre,twocolumn,groupedaddress,showpacs,superscriptaddress,amssymb,amsmath]{revtex4-1}
\usepackage[utf8]{inputenc}
\usepackage{graphicx}
\usepackage{tabularx}
\usepackage{xcolor}
\usepackage{amsmath}

\DeclareMathOperator{\csch}{csch}
\usepackage{comment}

\usepackage{dcolumn}
\usepackage{hyperref}
\hypersetup{
    colorlinks=true,
    linkcolor=blue,
	citecolor=cyan
}
\usepackage{bm}
\usepackage{epsf}
\usepackage{subfigure}
\usepackage{braket}
\usepackage{tensor}
\usepackage{soul}
\usepackage{float}

\usepackage{mathrsfs}
\usepackage{mathtools}
\usepackage{multirow}

\usepackage{amsfonts}

\DeclareMathOperator{\trace}{Tr}

\newcommand{\be}{\begin{equation}}
\newcommand{\ee}{\end{equation}}
\newcommand{\bea}{\begin{eqnarray}}
\newcommand{\eea}{\end{eqnarray}}

\newcommand{\mbar}[1]{\overline{#1}^{(\!N\!)}}
\newcommand{\mmbar}[1]{\overline{#1}^{(\!N+1\!)}}
\newcommand{\mzbar}[1]{\overline{#1}^{(0)}}

\def\l{\lambda}
\def\wl{\widetilde}
\def\b{\beta}

\def\la{\langle}

\def\ra{\rangle}

\def\om{\omega}
\def\dl{\Delta\lambda}
\def\db{\Delta\beta}
\def\gw{\gamma_w}
\def\gh{\gamma_h}
\def\e{\epsilon}
\def\h{\hbar}
\def\ve{\varepsilon}
\def\lf{\left}
\def\rt{\right}

\newcommand*\mathinhead[2]{\texorpdfstring{$\boldsymbol{#1}$}{#2}}

\makeatletter
\newsavebox{\@brx}
\newcommand{\llangle}[1][]{\savebox{\@brx}{\(\m@th{#1\langle}\)}%
  \mathopen{\copy\@brx\kern-0.5\wd\@brx\usebox{\@brx}}}
\newcommand{\rrangle}[1][]{\savebox{\@brx}{\(\m@th{#1\rangle}\)}%
  \mathclose{\copy\@brx\kern-0.5\wd\@brx\usebox{\@brx}}}
\makeatother

\makeatletter
\newcommand*\rel@kern[1]{\kern#1\dimexpr\macc@kerna}
\newcommand*\widebar[1]{%
  \begingroup
  \def\mathaccent##1##2{%
    \rel@kern{0.8}%
    \overline{\rel@kern{-0.8}\macc@nucleus\rel@kern{0.2}}%
    \rel@kern{-0.2}%
  }%
  \macc@depth\@ne
  \let\math@bgroup\@empty \let\math@egroup\macc@set@skewchar
  \mathsurround\z@ \frozen@everymath{\mathgroup\macc@group\relax}%
  \macc@set@skewchar\relax
  \let\mathaccentV\macc@nested@a
  \macc@nested@a\relax111{#1}%
  \endgroup
}
\makeatother

\begin{document}
\newcolumntype{M}[1]{>{\centering\arraybackslash}m{#1}}
\title{Full statistics of nonequilibrium heat and work for many-body quantum Otto engines and universal bounds: A nonequilibrium Green's function approach}
\author{Sandipan Mohanta}
\email{mohanta.sandipan@students.iiserpune.ac.in}
\affiliation{Department of Physics,
		Indian Institute of Science Education and Research, Pune 411008, India}
\author {Bijay Kumar Agarwalla}
\email{bijay@iiserpune.ac.in}
\affiliation{Department of Physics,
		Indian Institute of Science Education and Research, Pune 411008, India}
		\email{bijay@iiserpune.ac.in}
	
	\date{\today}
	
\begin{abstract}
We consider a generic four-stroke quantum Otto engine consisting of two unitary and two thermalization strokes with an arbitrary many-body working medium. Using the Schwinger-Keldysh nonequilibrium Green's function formalism, we provide an analytical expression for the cumulant generating function corresponding to the joint probability distribution of nonequilibrium work and heat. The obtained result is valid up to the second order of the external driving amplitude.    
We then focus on the linear response limit and obtain Onsager's transport coefficients for the generic Otto cycle and showed that the traditional  fluctuation-dissipation relation for the total work is violated in the quantum domain, whereas for heat it is preserved. This 
leads to remarkable consequences in obtaining universal constraints on heat and work fluctuations for engine and refrigerator regimes of the Otto cycle and further allows us to make connections to the thermodynamic uncertainty relations. These findings are illustrated using a paradigmatic model that can be feasibly implemented in experiments.


\end{abstract}
\maketitle


\section{Introduction}
Thermodynamic cycles are fundamental concepts in the field of thermodynamics that describe the exchange and conversion of energy and play a crucial role in understanding and optimizing the performance of various energy conversion devices \cite{cengel2004thermodynamics}. Traditionally, thermodynamics provides a description of systems in terms of bulk or macroscopic properties by focusing primarily on averages or mean values of thermodynamic quantities \cite{zemansky1968heat,callen-book}. However,  for small-scale systems consisting of a few degrees of freedom, thermal and quantum fluctuations about the mean values can be significant. Exploring the universal properties of these fluctuations for out-of-equilibrium systems has been of key interest in the field of stochastic and quantum thermodynamics \cite{fluc-3,st-thermo1, st-thermo2,Goold_2016, Q-thermo2}. Significant theoretical insight has been achieved in this regard over the past two decades, owing to the diverse kinds of universal fluctuation relations \cite{fluc-3, holubec2021fluctuations,fluc-1, fluc-2}.

In recent times, quantum engines have garnered significant interest as they provide a privileged platform  to investigate the intricate relationship between  quantum phenomena and nonequilibrium thermodynamics. Furthermore, they offer opportunities to harness the unique properties of the quantum working medium to boost performance and enhance stability \cite{quantum-supremacy,Vroylandt_2017,Niedenzu_2018,interaction-enhanced}. Thanks to rapid technological developments, it has now become possible to fabricate small-scale thermal machines  \cite{engine-expt-spin,myers2022quantum,blickle2012realization,martinez2016brownian,martinez2017colloidal,Eric-engine,engine-refg,ion-heat-engine,single-ion-eng,Zhang2022,Amit_Dutta-review}.  For example, the smallest possible quantum Otto engine, with a qubit as working fluid has been realized in an NMR setup  \cite{engine-expt-spin}. Another paradigmatic model, a harmonic oscillator heat engine has also been realized in the ion-trap setup \cite{single-ion-eng}. More complex and exotic many-body working fluids executing a cycle has now been proposed \cite{Rydberg_atom-eng,Chen2019,otto-Tavis_Cummings,otto-spin_squeezing,otto-ramandeep,otto-XX,Myers_2022,carnot-cycle-spin_system,baryton-cycle,Singh2020,quantum-supremacy}. 

The study of fluctuation in the context of such small-scale thermal machines is of tremendous significance \cite{watanabe2022finite, ito2019universal, Lutz1, Udo-TUR-bound, TUR_HO}. Recently, there have been intense ongoing efforts to quantify the performance of quantum engines involving efficiency, power, and power fluctuation or so-called constancy \cite{trade-off-engine}. In this regard, Thermodynamic uncertainty relations (TURs), establishing a trade-off relation between the relative fluctuation of thermodynamic currents and entropy production, have been put forward that rule out the possibility of achieving Carnot efficiency without allowing power fluctuation to diverge  \cite{Barato:2015:UncRel,trade-off-engine,Gingrich:2016:TUR,Falasco,Landi-Goold-TUR,multiterminal-TUR,Bijay-Dvira-TUR}.  Another direction of studies established universal upper and lower bounds on the ratio of power to input heat current fluctuations of generic continuously coupled  autonomous thermal machines operating in the linear response regime \cite{Universal-Agarwalla}. These studies have been further extended to discrete four-stroke Otto cycles with specific working mediums \cite{Sushant2021_otto,sandipan02} and very recently for model-specific periodically driven continuous machines \cite{ArpanDas2023}. However, a detailed understanding of the bounds on nonequilibrium fluctuations for non-autonomous machines with generic many-body working medium is still lacking. 

In this work, we consider a generic four-stroke quantum Otto cycle \cite{Campisi-PRB-2020,Otto-01,Otto-02,Otto-03,Otto-04,Otto-05,Otto-06,Otto-07,Otto-08,Otto-09,Otto-10,otto-kos-1,kos-2} consisting of an arbitrary many-body working medium and derive an analytical expression for the joint cumulant generating function (CGF) of total work and input heat, valid up to the second order of the driving amplitude, by employing the rigorous Schwinger-Keldysh nonequilibrium Green's function (NEGF) formalism \cite{NEGF-work,negf_massi, Rammer1, SchwingerNEGF}. The  CGF obtained for an arbitrary many-body working medium satisfies the nonequilibrium fluctuation relation \cite{Campisi_2014}. Furthermore, we demonstrate the linear response limit characterized by a small driving amplitude and small temperature difference and obtain the expressions for the Onsager's transport coefficients. Notably, we reveal a breakdown of the traditional fluctuation-dissipation relation  (FDR) \cite{kubo1966fluctuation,weber-FDT,Felderhof_1978,pathria2021statistical} for the total work, while, the traditional FDR remains intact for heat. Our findings have remarkable implications for the generic Otto cycle: when operating as an engine, the ratio of output (work) to input (heat from the hot reservoir) fluctuations is universally bounded from below, while in the refrigerator regime, a similar quantity is universally bounded from above. Additionally, we also establish connections to the TURs. Finally, we compare the obtained universal bounds for the discrete Otto cycle with those of autonomous steady-state machines \cite{Universal-Agarwalla,sandipan02,Sushant2021_otto,ArpanDas2023}.

We organize the paper as follows: Section \ref{sec:QOE} briefly describes the quantum Otto cycle and the two-point measurement protocol to construct the CGF corresponding to the joint probability distribution of total work and heat. We then demonstrate how to map the CGF on the modified Keldysh contour. In Section \ref{sec:pertur-exp}, we obtain the expression for the CGF valid up to the second order of the external driving amplitude. We validate that the CGF respects the fluctuation symmetry. In Section \ref{sec:linear}, we discuss the linear response limit, obtain the Onsager transport coefficients, and derive universal bounds on nonequilibrium fluctuations. We illustrate the obtained results with a paradigmatic model in Section \ref{sec:result}. Finally, we summarize our main findings and provide an outlook in Section \ref{sec:summary}.

\section{Quantum Otto cycle and Characteristic function on modified Schwinger-Keldysh contour}
\label{sec:QOE}
We consider a generic many-body working medium for the quantum Otto cycle and treat the many-body interaction part as a time-dependent perturbation. The total Hamiltonian of the working fluid is given by 
\begin{equation}
    H(t)=H_0+\l(t)H_1,
\end{equation}
where $H_0$ is the non-interacting part, often referred to here as the free Hamiltonian, and $H_1$ represents an arbitrary many-body interaction term. $\lambda(t)$ is an arbitrary driving protocol that drives the system away from equilibrium. Importantly, $\l(t)$  serves the purpose of a perturbation parameter, and the values it can take are significantly smaller than those of other dimensionally comparable free parameters. 
\begin{figure}
    \centering
     \includegraphics[width=1.0\columnwidth]{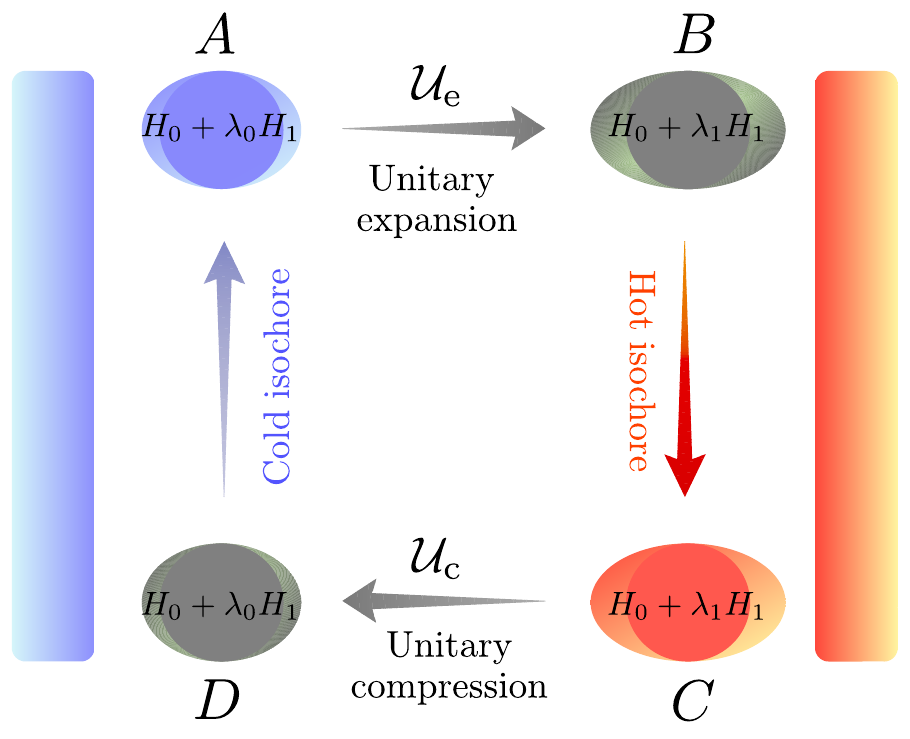}
    \caption{Schematic of a four-stroke Otto cycle with a many-body quantum working medium executing two unitary and two thermalization strokes. The expansion stroke is carried out by the unitary protocol ${\cal U}_{\rm e}$ and the compression stroke is carried out by the ${\cal U}_{\rm c}\!=\!\Theta\mathcal{U}_\mathrm{e}^\dagger\Theta^\dagger$. The final states of the unitary strokes at $B$ and $D$ are non-thermal ones. The isochoric thermalization strokes take place in the presence of the reservoirs.}
    \label{figschem}
\end{figure}
The working medium performs a four-stroke Otto cycle consisting of two unitary and two thermalization strokes with respect to the cold and hot thermal reservoirs with fixed inverse temperatures $\b_c$ and $\b_h$, respectively (see Fig.~\ref{figschem}). Below we briefly describe these four strokes:
(1) {\it Unitary expansion stroke} ($A \!\to\! B$). The system starts at time $t\!=\!0$ from the initial thermal state $A$  characterized by the cold inverse temperature $\beta_c$ and Hamiltonian $H(0)\!=\! H_0 + \l(0) H_1$, where $\l(0)\!=\!\l_0$. The system is then decoupled from the reservoir and $\l(t)$ drives the system out of equilibrium. The first unitary stroke ends at time $t\!=\!\tau_1$ with  $\l(\tau_1)\!=\! \l_1$. At the end of this stroke, the state of the system $B$ is a non-thermal one. (2) {\it Hot isochore} ($B\!\to\! C$). During the first heat exchange stroke,  the system is brought into weak contact with the hot reservoir while keeping $\l_1$ fixed. The system then waits until it reaches a thermal state $C$,  parameterized by ($\l_1,\b_h$) at time $t\!=\!\tau_1+\tau_h$. (3) {\it Unitary compression} ($C\!\to\! D$). The system is then decoupled from the hot reservoir and the interaction strength $\l(t)$ is decreased unitarily from $\l_1$ to $\l_0$, and consequently, the system ends up in a non-thermal state $D$ at time $t\!=\!\tau_1+\tau_h+\tau_2$. (4) {\it Cold isochore} ($D\!\to\!A$): In the final step, the system is placed in weak contact with the cold reservoir while keeping $\l_0$ fixed until it reaches the thermal state parametrized by ($\l_0,\b_c$) at time $t=\tau_1+\tau_h+\tau_2+\tau_c=\tau_{cyc}$. Here, we assume that the system achieves full thermalization during both hot and cold isochores. This assumption is valid when the heat exchange stroke times $\tau_h$ and $\tau_c$ are much larger than the system relaxation time. Additionally, we only consider here the symmetric driving case, meaning the unitary evolution operators, represented by $\mathcal{U}_\mathrm{e}$ and $\mathcal{U}_\mathrm{c}$, governing the expansion and compression strokes,  respectively,  are mirror images of each other, \emph{i.e.}, $\mathcal{U}_\mathrm{c}\!=\!\Theta \mathcal{U}_\mathrm{e}^\dagger\Theta^\dagger$, where $\Theta$ is the anti-unitary time-reversal operator, and as a result, the two unitary stroke duration are equal,  $\tau_1\!=\!\tau_2$, which we set as $\tau$.

Under the complete thermalization assumption, we construct the joint probability distribution (PD) of total work and heat exchange in the hot isochore $P(w,q_h)$ by employing the projective measurements of the energy of the system at the end points of each stroke i.e., at $A$, $B$, $C$, and $D$, respectively  \cite{Lutz1}. 
The characteristic function (CF) corresponding to the joint distribution is given by the Fourier transform of the PD,
\begin{align}
    \chi(&\gamma_w ,\gamma_h)=\int\!\!\!\int d{w}\,d{q_h}\, P({w},{q_h})\,e^{i\gamma_w{w}}\,e^{i\gamma_h{q_h}}\nonumber\\
    &=\frac{\mathrm{Tr}\big[{\cal U}_\mathrm{e}^\dagger \,e^{(i\gamma_w-i\gamma_h)H(\tau)}\,{\cal U}_\mathrm{e}\,e^{(-i\gamma_w-\beta_c) H(0)}\big]}{{\cal Z}_c[H(0)]}\nonumber\\&\times\frac{\mathrm{Tr}\big[{\cal U}_\mathrm{c}^\dagger\, e^{i\gamma_w H(0)}\,{\cal U}_\mathrm{c}\,e^{(-i\gamma_w+i\gamma_h-\beta_h) H(\tau)}\big]}{{\cal Z}_h[H(\tau)]},\label{CF}
\end{align}
where $\gamma_w, \gamma_h$ are the counting variables associated to the total work $w\!=\!w_1 + w_3$ and the heat exchange with the hot reservoir $q_h$, respectively. In Eq.~\eqref{CF}, $H(0)\!=\!H_0+\l_0 H_1$ and $H(\tau)\!=\!H_0+\l_1 H_1$. ${\cal Z}_c[H(0)]\!=\!\trace[e^{-\b_c H(0)}]$ and ${\cal Z}_h[H(\tau)]\!=\!\trace[e^{-\b_h H(\tau)}]$ are the canonical partition functions with respect to the cold and hot inverse temperatures $\b_c$ and $\b_h$, and Hamiltonians $H(0)$ and $H(\tau)$, respectively. We observe that, due to the complete thermalization assumption, the joint PD in Eq.~\eqref{CF} takes a product form. Furthermore, an intriguing implication of this assumption is that the joint PD of total work and heat from the hot reservoir alone suffices to determine the statistics of heat exchange with the cold reservoir  \cite{sandipan02}.

\begin{figure*}
    \centering
    \includegraphics[width=2.0\columnwidth]{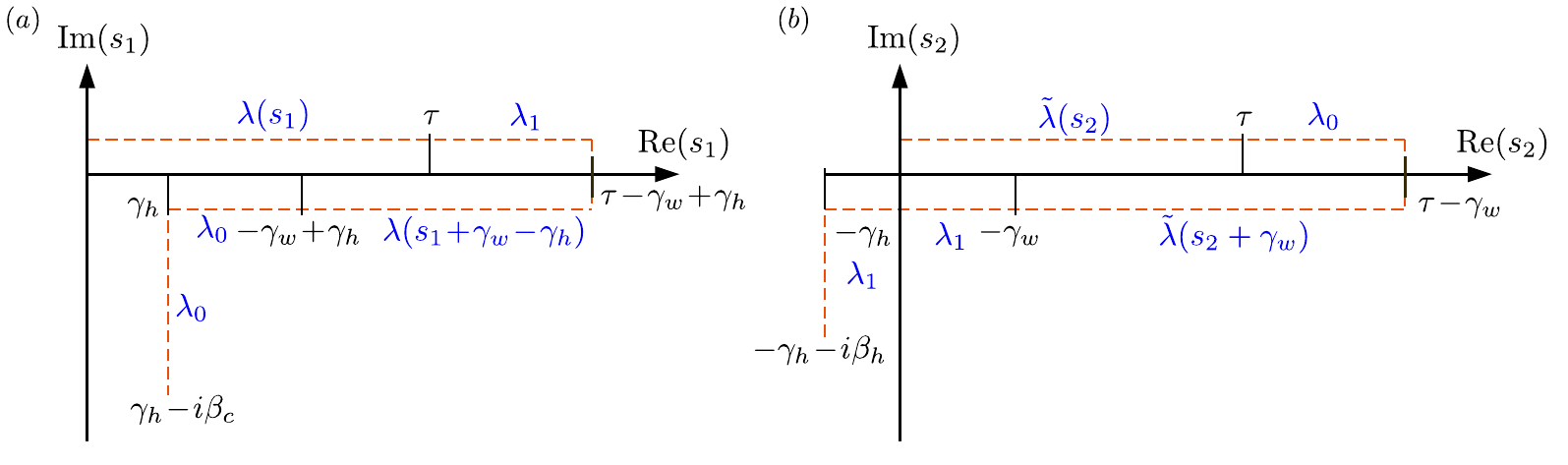}
    \caption{The modified Keldysh contour to compute joint work and heat statistics for a generic quantum Otto cycle. $(a)$ Contour $C$ corresponds to the numerator of the first term in Eq.~\eqref{CF}. Note that, the denominator of the first term runs only on the vertical tail as indicated by the integration limits. $(b)$ Contour $C'$ corresponds to the numerator of the second term in Eq.~\eqref{CF}. Here, we are considering symmetric driving only, \emph{i.e.}, $\mathcal{U}_c=\Theta^\dagger\mathcal{U}_e^\dagger\Theta$ [or in other words $\tilde{\l}(s)\!=\!\l(\tau\!-\!s)$]. The denominator of the second term runs only on the vertical tail as indicated by the integration limits. As mentioned in the main text, none of the modified contours $C$ and $C'$ return to $\mathrm{Re}(s)\!=\!0$ and get closed because of the presence of the non-unitary heat exchange stroke.}  
    \label{figcontour}
\end{figure*}

In order to obtain a perturbative expression of the CF
given in Eq.~\eqref{CF}, we utilize the Schwinger-Keldysh formal-
ism, a systematic approach for calculating
nonequilibrium Green’s functions and correlation functions under the influence of time-dependent perturbations by casting the problem on a closed time-ordered contour in the complex time plane, known as
the Keldysh contour. For the problem considered here,
all the unitary evolution operators  and the exponential operators, \emph{e.g.,} $e^{(i\gamma_w-i\gamma_h)H(\tau)}$, appearing in Eq.~\eqref{CF} are mapped on two modified Keldysh contours. The first trace in Eq.~\eqref{CF} corresponds to the modified contour $C$, as shown in Fig.~\ref{figcontour}(a), and the second trace gives rise to another modified contour $C'$, as shown in  Fig.~\ref{figcontour}(b). It is important to mention that, in the Schwinger-Keldysh formalism one typically expects closed contours while dealing with time-dependent perturbations. However, here, none of our modified contours are closed, only articulating the fact that both the traces that are mapped contain partial information about the non-unitary heat exchange stroke. The tails represent the presence of interaction $H_1$ in the thermal states at $A$ and $C$ (see Fig.~\ref{figschem}). We have demonstrated the details of this mapping in Appendix \ref{AppA} and here we only provide a compact expression for the joint CF on the Keldysh contour, 
\begin{align}
   \chi(\gamma_w,\gamma_h)&= \frac{\big\langle {\cal T}_C \, e^{-\frac{i}{\hbar}\int_C\! \lambda_C(s_1)H_1^I(s_1) ds_1}\big\rangle_{\wl{\b}_c}}{\big\langle {\cal T}_C \,e^{-\frac{i}{\hbar}\int_{\hbar\gamma_h}^{\hbar\gamma_h\!-\!i\hbar\beta_c}\! \lambda_0 H_1^I(s_1) ds_1}\big\rangle_{{\b}_c}}\nonumber\\&\times \,\frac{\big\langle {\cal T}_{C'} \,e^{-\frac{i}{\hbar}\int_C\! {\lambda}_{C'}(s_2)H_1^I(s_2)ds_2}\big\rangle_{\wl{\b}_h}}{\big\langle {\cal T}_{C'}\, e^{-\frac{i}{\hbar}\int_{\tau\!-\!\hbar\gamma_h}^{\tau\!-\!\hbar\gamma_h\!-\!i\hbar\beta_h}\! \lambda_1H_1^I(s_2) ds_2}\big\rangle_{{\b}_h}}\nonumber\\&\times \,\langle e^{-i\gamma_h H_0}\rangle_{\b_c}
   \langle e^{i\gamma_h H_0}\rangle_{\b_h}.
\label{contourCF}
\end{align}
Here, crucially, the averages in both  the numerators are taken with respect to shifted inverse temperatures $\wl{\b}_c\!=\!\b_c+i\gh$ and $\wl{\b}_h\!=\!{\b}_h-i\gamma_h$ which are counting-field dependent. ${\cal T}_C$ and ${\cal T}_{C'}$ are the contour-time ordered operators defined on the contours $C$ and $C'$, respectively, which orders the time-dependent operators  according to their contour-time argument; \emph{i.e.}, sorting the operators from left to right with their contour-time arguments
decreasing. In Appendix \ref{AppB} we have expanded the exponential operators in Eq.~\eqref{contourCF} up to the second order in driving protocol $\l(s)$ and obtained the expression of the joint cumulant generating function (CGF) $\ln{\chi(\gw,\gh)}$ for arbitrary many-body working medium. The CGF is a useful quantity as it enables us to calculate all the cumulants (denoted by double angular brackets) of total work and heat exchange with the hot reservoir by taking partial derivatives with respect to $\gamma_w$ and $\gamma_h$, respectively,
\begin{align}
    \llangle w^k q_h^l \rrangle = \frac{\partial^k\partial^l}{\partial(i\gamma_w)^k\partial(i\gamma_h)^l}\,\ln{\chi(\gamma_w,\gamma_h)}\Big|_{\gamma_w,\gamma_h = 0}.
\end{align}

\section{Work and heat cumulants for arbitrary many-body working medium}
\label{sec:pertur-exp}
In this section, we write down the final expression for the CGF valid up to the second order of the driving protocol $\l(s)$  (see Appendix \ref{AppB} for the details). We obtain  
\begin{widetext}
\begin{align}
     \ln \chi(&\gamma_w,\gamma_h)\!=\!\big[\!-\!i\gamma_w(\lambda_1\!-\!\lambda_0)+i\gamma_h\lambda_1\big]\,\big[\langle H_1\rangle_{\wl{\beta}_h}
     \!\!-\!\langle H_1\rangle_{\wl{\beta}_c}\big]+ \big[\!-\!i\h\gamma_w(\lambda_1^2\!-\!\lambda_0^2)+i\h\gamma_h\lambda_1^2\big]\! \int\frac{d\omega}{2\pi}\frac{\wl{G}_h^>(\omega)\!-\!\wl{G}_c^>(\omega)}{\omega}\nonumber\\
     &-\!\beta_c\lambda_0\Big[\langle H_1\rangle_{\wl{\beta}_c}\!\!-\!\langle H_1\rangle_{\beta_c}+\h\lambda_0\!\int\frac{d\omega}{2\pi}\frac{\wl{G}_c^>(\omega)\!-\!{G}_c^>(\omega)}{\omega}\Big]
     \!-\!\beta_h\lambda_1\Big[\langle H_1\rangle_{\wl{\beta}_h}\!\!-\!\langle H_1\rangle_{\beta_h}+\h\lambda_1\!\int\frac{d\omega}{2\pi}\frac{\wl{G}_h^>(\omega)\!-\!{G}_h^>(\omega)}{\omega}\Big]\nonumber\\
    &+\!\int\frac{d\omega}{2\pi}\frac{1\!-\!e^{i\h\omega(\gamma_w\!-\!\gamma_h)}}{\omega^2}A(\omega)\wl{G}_c^>(\omega)
     +\!\int\frac{d\omega}{2\pi}\frac{1\!-\!e^{i\h\omega\gamma_w}}{\omega^2}{A}(\omega)\wl{G}_h^>(\omega)+\ln{\langle e^{-i\gamma_h H_0}\rangle_{\beta_c}}+\ln {\langle e^{i\gamma_h H_0}\rangle_{\beta_h}} .
     \label{main-CGF}
\end{align}
\end{widetext}
Equation~\eqref{main-CGF} is the first central result of this paper. It describes the statistical properties of total work and input heat fluctuations in the many-body Otto cycle in terms of one-point and two-point correlators of the interaction Hamiltonian $H_1$. In this expression, we suppress the integration limits for $\omega$, which ranges from $-\infty$ to $+\infty$. Here, the nonequilibrium Green's functions are defined as connected two-point correlators of $H_1$ in the interaction picture,  
\begin{align}
 \wl{G}_j^>(s)&=\left(\!-\frac{i}{\hbar}\right)^{2} \!\!\llangle H_1^I(s)H_1\rrangle_{\wl{\beta}_j}, \label{gt>}\\
 {G}_j^>(s)&=\left(\!-\frac{i}{\hbar}\right)^2 \!\!\llangle H_1^I(s)H_1\rrangle_{{\beta}_j},
 \label{g>}
\end{align}
where $j\!=\!c,h$. In this context, $\llangle \cdot \rrangle$ implies a connected correlation function, also known as a cumulant correlation function,
\begin{equation}
    \llangle X^I(s_1) Y^I(s_2) \rrangle_\b = \langle X^I(s_1) Y^I(s_2) \rangle_\b \!-\! \langle X \rangle_\b \langle Y \rangle_\b. \label{llangle}
\end{equation} 
The shifted inverse temperatures $\wl{\beta}_j$, $j\!=\!c,h$, appearing in Eq.~\eqref{main-CGF} play a crucial role in calculating the cumulants of the heat exchange.
In this paper, we adopt  the Fourier transformation convention as
\begin{equation}
     {G}_j^>(\omega)=\!\int_{-\infty}^{+\infty}\!\!ds\,{G}_j^>(s) \, e^{i\omega s}.
\end{equation}   
Crucially, all the information about the time-dependent driving protocol is encapsulated in the quantity $A(\om)$, defined as
\begin{align} A(\om)=\left|\int_0^\tau\!\dot{\l}(t)e^{i\om t} d t\right|^2.
\end{align}
It is worth nothing that the CGF in Eq.~\eqref{main-CGF} satisfies the normalization condition, $ \ln{\chi(0,0)}\!=\!0$. More importantly, it satisfies the universal  heat engine  fluctuation symmetry \cite{Campisi_2014,Campisi2015,JWfluc},
\begin{equation}
    \ln \chi\big[ i\beta_c, \, i(\beta_c\!-\!\beta_h)\big] =0.\label{ft}
\end{equation}
Notably, this fluctuation symmetry is a direct consequence of the Kubo-Martin-Schwinger (KMS) boundary condition satisfied by the Green's functions introduced in Eqs.~\eqref{gt>} and \eqref{g>} \cite{breuer2002,carmichael1993open,kubo1957,martin-schwinger},
\begin{align}
    &\wl{G}^>(s\!-\!i\h\wl{\b}_c)=\wl{G}^>(-s),\label{kms1}\\
    &{G}^>(s\!-\!i\h{\b}_c)={G}^>(-s).\label{kms2}
\end{align}
Here, $\wl{\beta}_c$ and ${\beta}_c$ are the shifted and original inverse temperatures of the cold reservoir, respectively. Furthermore, the fluctuation symmetry in Eq.~\eqref{ft} implies that  $\langle e^{-\Sigma} \rangle \!=\!1$, where $\Sigma\!=\!\b_c w+(\b_c\!-\!\b_h) q_h $ is the stochastic entropy production in a single cycle. It is worth highlighting that utilizing Jensen's inequality, $\la e^{-\Sigma}\ra\ge e^{-\la \Sigma\ra}$, we immediately recover the second law of thermodynamics, $\la \Sigma\ra\ge0$ \cite{landi-irreversible-entropy}.

In the following, we present the formal expressions of the first and second cumulants of total work and heat exchange with the hot reservoir. The average total work is given by
\begin{align}
     \langle w\rangle \!=&-\! (\lambda_1\!-\!\lambda_0)\lf[\langle H_1\rangle_{\beta_h}\!\!-\!\langle H_1\rangle_{\beta_c}\rt]\nonumber\\&-\h(\lambda_1^2\!-\!\lambda_0^2)\!\int\!\frac{d\omega}{2\pi}\frac{1}{\omega}\left[{G}_h^>(\omega) \!-\!{G}_c^>(\omega)\right]\nonumber\\&-\!\h\!\int\!\frac{d\omega}{2\pi}\frac{A(\omega)}{\omega}\left[G^>_c(\omega)+G^>_h(\omega)\right].
     \label{w-avg}
     \end{align}
The first two terms on the right-hand side represent the boundary terms that depend on the end values of the driving protocol and are associated with the total change in free energy during the two unitary work strokes.

The corresponding variance (second cumulant) of total work is given by 
\begin{equation}     
     \llangle w^2\rrangle =-\h^2\!\int\!\frac{d\omega}{2\pi}A(\omega)\lf[G^>_c(\omega)+G^>_h(\omega)\rt].
     \label{w-fluc}
\end{equation}
Notably, we find that calculating the cumulants of heat exchange $q_h$ requires more involved calculations due to the presence of shifted inverse temperatures $\wl{\b}_c$ and  $\wl{\b}_h$. This results in connected correlators of the form $\llangle H_1H_0\rrangle$, $\llangle H_1^I(s) H_1 H_0 \rrangle$, and so on. We note that, for an arbitrary operator $X$, 
\begin{gather}
    \frac{d}{d(i\gamma_h)}\langle X\rangle_{\wl{\b}_j}\!=\!\mp\llangle X H_0\rrangle_{\wl{\b}_j},\\
     \frac{d^2}{d(i\gamma_h)^2}\la X\ra_{\wl{\b}_j}\!=\!\llangle X H_0^2\rrangle_{\wl{\b}_j},
\end{gather}
where $(\!-\!)$ corresponds to  $j\!=\!c$, and $(+)$ corresponds to $j\!=\!h$, respectively.
Below we list the expressions for the average heat exchange with the hot reservoir and its variance,
\begin{widetext}
\begin{align}
         \langle q_h\rangle \!=& \langle H_0\rangle_{\b_h}\!\!-\!\langle H_0\rangle_{\b_c}+ \lambda_1\big[\langle H_1\rangle_{\b_h}\!\!-\!\langle H_1\rangle_{\b_c}\big]+\h\lambda_1^2\!\int\!{\frac{d\omega}{2\pi}\frac{{G}_h^>(\omega)\!-\!{G}_c^>(\omega)}{\omega}}+\beta_c \lambda_0\Big[\llangle H_0H_1\rrangle_{\b_c}\!+\h\lambda_0 \!\int\! \frac{d\omega}{2\pi}\frac{G^{'}_c(\omega)}{\omega}\Big]  \nonumber \\
         &\quad\quad \quad\quad-\!\beta_h \lambda_1\Big[\llangle H_0H_1\rrangle_{\b_h}\!+\h\lambda_1\! \int\! \frac{d\omega}{2\pi}\frac{G^{'}_h(\omega)}{\omega}\Big]+\h\!\int\!\frac{d\omega}{2\pi}\frac{A(\omega)G^>_c(\omega)}{\omega},
         \label{q-avg}\\
    \llangle q_h^2\rrangle \!=& \llangle H_0^2\rrangle_{\b_c}\!+\llangle H_0^2\rrangle_{\b_h}\!+2\l_1\big[\llangle H_0 H_1\rrangle_{\b_c}\!+\llangle H_0 H_1\rrangle_{\b_h}\big]+2\h\l_1^2\int\!\frac{d\om}{\pi}\frac{G_c^{'}(\om)+G_h^{'}(\om)}{\om}\!-\!\b_c\l_0\Big[\llangle H_0^2 H_1\rrangle_{\b_c}\nonumber\\
   & \!\!+\h\l_0\!\int\!\frac{d\om}{2\pi}\frac{G_c^{''}(\om)}{\om}\Big]\!-\!\b_h\l_1\Big[\llangle H_0^2 H_1\rrangle_{\b_h}\!+\h\l_1\!\int\!\frac{d\om}{2\pi}\frac{G_h^{''}(\om)}{\om}\Big]+2\h\!\int\!\frac{d\om}{2\pi}\frac{G^{'}_c(\om)A(\om)}{\om}+\h^2\!\int\!\frac{d\om}{2\pi}{A(\om) G_c^>(\om)},
   \label{q-fluc}
\end{align}
\end{widetext}
where we have introduced a compact notation,
\begin{eqnarray}
    G_j^{'}(\omega) &=&-\frac{\partial}{\partial\beta_j}G_j^>(\omega), \\
    G_j^{''}(\omega)&=&\frac{\partial^2}{\partial\beta_j^2}G_j^>(\omega),
\end{eqnarray}
with $j=c,h$.

Now that we have obtained  the expressions for
the mean and the variance for total work and heat exchange with the hot reservoir from the joint CGF, in the following section, we focus on establishing the linear response regime for the generic quantum Otto cycle. We derive the expressions for Onsager's transport coefficients and explore the fate of traditional fluctuation-dissipation relations for work and heat.

\section{Linear response, Onsager coefficients, and universal bounds}
\label{sec:linear}
Let us first  parameterize the driving protocol $\l(t)$ as
\begin{align}
    \l(t)=\l+\dl \, f(t),
    \label{nFt}
\end{align}
where, $\l\!=\!\l_0$ and $\dl\!=\!\l_1 \!-\!\l_0$. The function $f(t)$ is defined such that $f(0)\!=\!0$ and $f(\tau)\!=\!1$ to ensure that $\lambda_0$ and $\lambda_1$ are the boundary values of the protocol $\lambda(t)$ at time $t\!=\!0$ and time $t\!=\!\tau$, respectively. Recall that $\tau$ is the duration of each work stroke. To identify the proper thermodynamic affinities and the corresponding conjugate fluxes, we begin with the definition of the stochastic entropy production in a single cycle:
\begin{align}
    \Sigma=&\b_c w+(\b_c\!-\!\b_h)\, q_h\nonumber\\
    =&\bm{(}\b_c \dl\bm{)}\frac{w}{\dl}+\bm{(}\b_c\!-\!\b_h\bm{)}q_h. 
\end{align}
From this expression, we can identify the thermodynamic affinities as ${\cal F}_w\!=\!\b_c\dl$ and ${\cal F}_q\!=\!\b_c\!-\!\b_h$, and the corresponding integrated work and heat fluxes as $w/\dl$ and $q_h$, respectively. One can further define the heat and work currents by dividing the integrated heat and work fluxes with total cycle time  $\tau_{cyc}\!=\!2\tau+\tau_h+\tau_c$, as follows:
\begin{gather}
   \;\;\mathcal{J}_w\!=\!\frac{1}{\tau_{cyc}}\frac{w}{\dl}  \;\;\mathrm{and}  \;\; \mathcal{J}_h\!=\!\frac{q_h}{\tau_{cyc}}.
\end{gather}
Consequently, the average entropy production rate in one complete cycle  $\la\sigma\ra\!=\!\la\Sigma\ra/\tau_{cyc}$ can be expressed as
\begin{align}
    \la\sigma\ra=\sum_i\la\mathcal{J}_i\ra\mathcal{F}_i,
\end{align}
where $i\!=\!w,h$. From this point onwards, we adopt $\b_c\!=\!\b$ and $\beta_c\!-\!\b_h\!=\!\Delta\b$, and focus on the linear response regime where $\db\ll\b$.  In this regime, we can effectively utilize the following expansion,
\begin{align}
    \la X\ra_{\b\!-\!\db}=&\la X\ra_{\b}\!-\!\db\,\frac{\partial}{\partial\b}\la X\ra_\b\nonumber\\
    =& \la X\ra_{\b}+\db\llangle X H_0\rrangle_{\b}.\label{taylor-expand}
\end{align}
Note that the appearance of the free Hamiltonian $H_0$ in the second term arises from the derivative with respect to $\beta$ over a canonical distribution characterized by $e^{-\beta H_0}/\trace{[e^{-\beta H_0}]}$. To proceed with our linear response analysis for the generic Otto cycle, we calculate the average work and heat fluxes, $\la \mathcal{J}_w\ra$ and $\la \mathcal{J}_h\ra$, using Eqs.~\eqref{w-avg} and \eqref{q-avg}, respectively, which are accurate up to the second order of the driving protocol $\l(t)$. Finally, carrying out the Taylor expansion given in Eq.~\eqref{taylor-expand} and keeping terms up to the linear order of the affinities, $\mathcal{F}_w$ and $\mathcal{F}_q$, we derive the following relation between the currents and affinities, connected via the Onsager matrix:
\begin{align}
    \begin{pmatrix}
    \la\mathcal{J}_w\ra\\
    \la\mathcal{J}_h\ra
    \end{pmatrix}
    =\begin{pmatrix}
    {\cal L}_{w w} & {\cal L}_{w q}\\
    {\cal L}_{q w} & {\cal L}_{q q}
    \end{pmatrix}
    \begin{pmatrix}
    {\cal F}_w\\
    {\cal F}_q
    \end{pmatrix},\label{onsagar}
\end{align}
where  the Onsager transport coefficients ${\cal L}_{ij}$, with $i,j=w, q$, are given by 
\begin{align}
    {\cal L}_{w w}\!&=\!-\frac{1}{\tau_{cyc}}\h\!\int\!\frac{d\om}{2\pi}\frac{2 F(\om)G^>(\om)}{\b\, \om},\label{Lww}\\
    {\cal L}_{w q}\!&=\!{\cal L}_{q w}\!=\!\frac{1}{\tau_{cyc}}\frac{\partial}{\partial\b}\Big[\la H_1\ra_\b\!+\!2\h\l\!\!\int\!\frac{d\om}{2\pi}\frac{G^>(\om)}{\om}\Big],\label{Lwq}\\
    \mathcal{L}_{q q}\!&=\!-\frac{\partial}{\partial\beta}\Big[\langle H_0\rangle_\beta\! +\!\lambda\frac{\partial }{\partial \beta}\Big(\beta\langle H_1\rangle_\beta\!+ \hbar\lambda\beta\!\!\int\!\frac{d\omega}{2\pi}\frac{G^>(\omega)}{\omega}\Big)\Big]\label{Lqq}.
    \end{align}
Here, note that 
\begin{align}
        F(\om)=\lf|\int_0^\tau\!\dot{f}(t)e^{i\om t} d t\rt|^2\!=\frac{A(\om)}{{\dl}^2}
\end{align}
encodes the information of driving where $\dl$ and $f(t)$ are defined in Eq.~\eqref{nFt}. The Green's function $G^>(\omega)$ is evaluated at the equilibrium temperature $\beta$. Importantly, we recover the Onsager reciprocity relation \cite{onsager1931}, \emph{i.e.} $\mathcal{L}_{wq}\!=\!\mathcal{L}_{qw}$, as shown in Eq.~\eqref{Lwq}. The derived expressions for the Onsager transport coefficients, characterizing the behavior of an arbitrary working medium executing the Otto cycle in the linear response regime, constitute another central result of our work. With these results in hand, we proceed to provide some important insights and remarks.

The traditional ``Fluctuation-Dissipation relations" (FDRs) typically connect the diagonal elements of the Onsager matrix to equilibrium fluctuations \cite{kubo1966fluctuation}. Remarkably, in our analysis, we precisely recover the traditional FDR for the heat current $\la \mathcal{J}_h\ra$ (detailed derivation in Appendix \ref{AppE}),
    \begin{align}
       {{\cal L}_{q q}}= \frac{\llangle\mathcal{J}_h^2\rrangle}{2}\bigg|_{\mathcal{F}_w, \mathcal{F}_q=0},
       \label{heat-FDR}
\end{align}
    where we recall that $\llangle\mathcal{J}_h^2\rrangle \!=\! \llangle q_h^2 \rrangle/\tau_{cyc}$. However, importantly, the same is not true for the case of work current. In fact, for the work current fluctuation, following Eq.~\eqref{w-fluc}, we obtain  
    \begin{align}
           \frac{\llangle\mathcal{J}_w^2\rrangle}{2}\bigg|_{\mathcal{F}_w, \mathcal{F}_q=0}=-\frac{1}{\tau_{cyc}}\h^2\!\int\!\frac{d\om}{2\pi} F(\om)G^>(\om),\label{varjw}
      \end{align}
      which does not match the diagonal Onsager coefficient $\mathcal{L}_{ww}$. To investigate this discrepancy further, let us rewrite the expressions for $\mathcal{L}_{ww}$ and $\llangle\mathcal{J}_w^2\rrangle$ as follows: 
      \begin{gather}
         {\cal L}_{w w}\!=\!-\frac{1}{\tau_{cyc}}\frac{\h^2}{2}\!\!\int\!\frac{d\om}{2\pi}F(\om)\frac{\tanh{\big(\frac{\b\h\om}{2}\big)} }{\frac{\b\h\om}{2}}\big[G^>(\om)+G^>(\!-\om)\big].
        \label{Lww1}\\
        \frac{\llangle\mathcal{J}_w^2\rrangle}{2}\bigg|_{\mathcal{F}_w, \mathcal{F}_q=0} \!=\!\!-\frac{1}{\tau_{cyc}}\frac{\h^2}{2}\!\int\!\frac{d\om}{2\pi} F(\om)\big[G^>(\om)+G^>(\!-\om)\big],\label{jw2-pre}
        \end{gather}
        Here, we have utilized an important identity involving the Green's functions,
\begin{equation}
 G^>(\om) \!-\! G^>(\!-\om)=\tanh{\Big(\frac{\b\h\om}{2}\Big)}\big[G^>(\om)+G^>(\!-\om)\big].
\end{equation}
which arises from the KMS condition in the Fourier space, $e^{-\b\h\om}G^>(\om)\!=\!G^>(-\om)$. Now, considering the fact that ${\tanh{(x)}}/{x}\!\le\!1$, we deduce from Eqs.~\eqref{jw2-pre} and \eqref{Lww1} that
    \begin{align}
       {\cal L}_{w w}\leq\frac{\llangle\mathcal{J}_w^2\rrangle}{2}\bigg|_{\mathcal{F}_w,\mathcal{F}_q=0} ,\label{w fdr}
    \end{align}
    highlighting a breakdown of the traditional FDR for work current.  Importantly, this breakdown is purely of quantum origin, and the equality is restored in the quantum-adiabatic limit, \emph{i.e.}, when $[H_0, H_1]\!=\!0$. This observation aligns with the recent findings of Ref.~\onlinecite{janet-workfluc, janet-workfluc2} where a similar breakdown was reported.

The obtained results for the Onsager coefficients and the FDR  breakdown for work current have remarkable implications in determining the universal bounds on the performance of the quantum Otto cycle operating as an engine or refrigerator. Particularly, it is noteworthy that the thermodynamic uncertainty relation \cite{Sacchi-TUR-1, Sacchi-TUR-2} holds for the currents  $\la \mathcal{J}_w \ra$ and $\la \mathcal{J}_h \ra$, as shown in Appendix \ref{AppD},  
\begin{align}
\langle \sigma \rangle \frac{\llangle\mathcal{J}_w^2\rrangle}{\langle \mathcal{J}_w \rangle^2} \geq 2, \quad 
\langle \sigma \rangle \frac{\llangle\mathcal{J}_h^2\rrangle}{\langle \mathcal{J}_h \rangle^2} \geq 2,
\end{align}
where $\langle \sigma \rangle = \langle \Sigma \rangle / \tau_{\textrm{cyc}}$ represents average entropy production rate. Notably, these  TURs remain valid independently of the nature of the Otto cycle's operation, \emph{i.e.}, whether it acts as an engine, refrigerator, heater, or accelerator \cite{Campisi-PRB-2020}. 

Let us now focus on understanding bounds on nonequilibrium fluctuations when the Otto cycle works as an engine, \emph{i.e.}, converting a fraction of spontaneous heat flux flowing from hot to cold reservoir  into useful work. To quantify the engine's performance, we introduce the power output as 
\begin{equation}
    \la \mathcal{P} \ra= -\la w \ra/\tau_{cyc} = \!-\frac{{\cal F}_w}{\b} \la \mathcal{J}_w\ra
\end{equation} 
and the input heat current is given by $\la \mathcal{J}_h  \ra$. The engine operational regime of the Otto cycle is characterized by $\la \mathcal{P}\ra\!>\!0$, and $\la \mathcal{J}_h\ra\!>\!0$, as per our sign convention where energy flowing into the working fluid is considered positive. To achieve this, the necessary requirement  is $\mathcal{F}_q\!>\!\mathcal{F}_w$.  Following the second law of thermodynamics, $\la \sigma \ra \!\ge\!0$, it is straightforward to show that the average efficiency of the Otto engine, $\la \eta \ra \!\coloneqq\! \la \mathcal{P} \ra / \la \mathcal{J}_h \ra$, is universally upper bounded by the the Carnot bound $\eta_c\!=\!\Delta \b/ \b$ \cite{callen-book,Carnot-2}, $\la \eta\ra\!\le\!\eta_c$.  In order to derive bounds on nonequilibrium fluctuations, following Ref.~\cite{Universal-Agarwalla}, we construct the quantity $\eta^{(2)}$, which is the ratio of power fluctuation to the fluctuation of input heat current,
\begin{align}
    \eta^{(2)}= \frac{\llangle\mathcal{P}^2\rrangle}{\llangle\mathcal{J}_h^2\rrangle}.
    \label{defeng}
\end{align}
Recently, it was shown that in the case of a continuously coupled  autonomous steady-state heat engine operating in the linear response regime, where both the work and heat FDRs are valid, the ratio of output power to input heat current fluctuations is subject to both universal lower and upper bounds \cite{Universal-Agarwalla}, 
\begin{align}
     \langle \eta \rangle^2 \leq   \big[\eta^{(2)}\big]_{\mathrm{auto}}  \leq \eta_c^2,
\end{align}
   where the subscript `$\mathrm{auto}$' stands for the autonomous  steady-state heat engine.
 However, in our case of a generic finite-time Otto engine operating in the linear response limit, we obtain the following inequality (see Appendix \ref{AppE} for details),
\begin{equation}
    \eta_c^2\ge \frac{{\cal F}_w^2{\cal L}_{w w}}{\b^2{\cal L}_{q q}}\ge \la \eta\ra^2.\label{eng-ons}
\end{equation}
A key point to note here is that, contrary to the autonomous steady-state engine, we observe a violation of the traditional work FDR in the discrete Otto cycle, as given in Eq.~\eqref{w fdr}. This FDR breakdown for the work current leads to the following inequality:
\begin{align}
    \eta^{(2)}\ge\frac{{\cal F}_w^2{\cal L}_{w w}}{\b^2{\cal L}_{q q}}.
    \label{def-eng}
\end{align}
Importantly, the equality is restored when the driving Hamiltonian $H_1$ commutes with the free Hamiltonian $H_0$, \emph{i.e.}, in the  quantum-adiabatic driving limit. As an immediate consequence of the inequality in Eq.~\eqref{def-eng}, the lower bound on $\eta^{(2)}$ remains robust, 
\begin{align}
    \eta^{(2)}\ge \la \eta\ra^2.\label{lower-bound}
\end{align}
However, the upper bound on $\eta^{(2)}$  may not always hold true, \emph{i.e.,} $\eta^{(2)} \nleq \eta_c^2$, the reason being purely of quantum origin.  This distinction is a significant departure from the autonomous case, where both the lower and upper bounds remain intact. In the Otto cycle, due to the violation of the traditional work FDR, the upper bound is not guaranteed, highlighting the impact of nonequilibrium and quantum effects on the engine's performance.
This constitutes another central result of this work.  As a direct consequence of the lower bound for $\eta^{(2)}$ in Eq.~\eqref{lower-bound}, a hierarchical relation between the TURs is observed. This follows from the fact that
\begin{equation}
\eta^{(2)}\!\ge\!\la \eta\ra^2\Rightarrow\frac{\llangle \mathcal{J}_w^2\rrangle}{\la \mathcal{J}_w\ra^2}\!\ge\!\frac{\llangle \mathcal{J}_h^2\rrangle}{\la \mathcal{J}_h\ra^2}.
\end{equation}
As a result, in the engine regime, a strict hierarchy between the TURs of work (output) and heat (input) currents is obtained:
\begin{align}
\langle \sigma \rangle \frac{\llangle\mathcal{J}_w^2\rrangle}{\langle \mathcal{J}_w \rangle^2} \geq 
\langle \sigma \rangle \frac{\llangle\mathcal{J}_h^2\rrangle}{\langle \mathcal{J}_h \rangle^2} \!\geq\! 2.\label{tur-eng}
\end{align}

After our exploration of the engine regime, we now shift our focus to the refrigerator regime.  In this mode of operation, the primary objective is to extract heat from the cold reservoir by utilizing external work. Following our sign convention, \emph{i.e.}, energy entering the working fluid is considered positive, the refrigerator regime is characterized by a positive current flowing out of the cold reservoir, $\la \mathcal{J}_c\ra\!>\!0$, and  work is performed on the system,  $\la\mathcal{J}_w\ra\!>\!0$. To achieve refrigeration, it is necessary that the affinities satisfy $\mathcal{F}_w\!>\!\mathcal{F}_q$. The performance of the refrigerator is quantified by its coefficient of performance (COP), denoted as $\la\varepsilon\ra$. Notably, the COP of a refrigerator is universally upper bounded by the Carnot COP  \cite{callen-book, Abah_2016}, \emph{i.e.,}
 \begin{align}
     \la\varepsilon\ra\coloneqq\frac{\b\la\mathcal{J}_c\ra}{\mathcal{F}_w\la\mathcal{J}_w\ra}\le \varepsilon_c,
 \end{align}
 where $\varepsilon_c\!=\!(1-\eta_c)/\eta_c\approx\b/\Delta\b$. 
Let us now investigate the quantity $\varepsilon^{(2)}$ defined as
\begin{align}
\varepsilon^{(2)}\coloneqq\frac{\b^2\llangle\mathcal{J}_c^2\rrangle}{\mathcal{F}_w^2\llangle\mathcal{J}_w^2\rrangle}.
\label{def-ref}
\end{align}
In Appendix \ref{AppF} we have derived the expression of CGF for the heat extracted from the cold reservoir,  from which we can obtain the expressions for the average heat extracted and its variance. It is worth mentioning that for steady-state autonomous refrigerators operating in the linear response regime, the quantity $\varepsilon^{(2)}$ was reported to possess both universal upper and lower bounds \cite{sandipan01}, 
\begin{equation}
    \ve_c^2\ge\big[\ve^{(2)}\big]_\mathrm{auto}\ge\la \ve\ra^2.
\end{equation}
where, recall that the subscript `$\mathrm{auto}$' denotes the autonomous steady-state refrigerator. In our case of a generic Otto cycle operating as a  refrigerator, we derived the following inequality (see Appendix \ref{AppG} for details):
\begin{equation}
    \ve_c^2\ge \frac{\b^2{\cal L}_{q q}}{{\cal F}_w^2{\cal L}_{w w}}\ge \la \ve\ra^2.\label{ref-ons}
\end{equation}
 Interestingly, the breakdown of the traditional work FDR leads to the following inequality for the quantity $\varepsilon^{(2)}$:
 \begin{align}
     \varepsilon^{(2)}\le\frac{\b^2{\cal L}_{qq}}{{\cal F}_w^2{\cal L}_{ww}},
 \end{align}
 where it is worth noting that the opposite inequality arises due to the presence of $\llangle \mathcal{J}_w^2 \rrangle$ in the denominator of the definition of $\varepsilon^{(2)}$.
 As a consequence, we observe a contrasting trend  compared to the engine regime. Specifically, in the refrigeration regime, the lower bound may not always be valid,   \emph{i.e.,} $\ve^{(2)}\!\ngeq\!\la\ve\ra^2$, whereas the upper bound remains intact,
 \begin{equation}
    \ve^{(2)}\!\leq\!\ve_c^2.
 \end{equation}
 Furthermore, it is essential to mention that the potential violation of the lower bound for $\ve^{(2)}$ leads to the loss of  the hierarchy in the TURs within the refrigerator regime, unlike what we observed in the engine regime of the Otto cycle as given in Eq.~\eqref{tur-eng}.  Intriguingly, a similar observation was recently reported for a specific model-dependent periodically driven continuous machine \cite{ArpanDas2023}. To summarize, the violation of the traditional FDR for the work current in the discrete Otto cycle gives rise to distinct and remarkable consequences for the engine and refrigerator regimes, thereby revealing the impact of quantum effects  on the discrete Otto cycle. It is only in the quantum-adiabatic driving limit, \emph{i.e.,} when $[H_0, H_1]\!=\!0$, that both the upper and lower bounds on $\eta^{(2)}$ and $\varepsilon^{(2)}$ are restored, akin to what is observed in  continuously coupled autonomous thermal machines.

\section{Example}
\label{sec:result}
To illustrate our findings, we now examine a specific model example. Our working medium consists of $N$ non-interacting bosons confined in a one-dimensional parabolic trap with frequency $\omega_0$. During the unitary strokes, the potential is perturbed with time. The unperturbed and the perturbing Hamiltonians of the working medium are given by 
\begin{align}
    H_0 &=\sum_{i=1}^{N}\frac{p_i^2}{2m} +\frac{1}{2}m \omega_0^2 x_i^2,\\
    H_1 &=\sum_{i=1}^{N}m\om_0^2 x_i^2,
\end{align}
respectively, such that $H(t)\!=\!H_0+\l(t) H_1,$ where, $\l(0)\!=\!\l$ and $\l(\tau)\!=\!\l+\Delta\l$. In the second-quantized formulation, we have
\begin{align}
    H_0&=\sum\nolimits_m \e_m a_m^\dagger a_m,\\
    H_1&=\frac{\hbar\om_0}{2}\sum\nolimits_m\sqrt{(m+1)(m+2)}\big(a_m^\dagger a_{m+2}+a_{m+2}^\dagger a_m\big)\nonumber\\&\quad\quad\quad\quad+(1+2m) a_m^\dagger a_m,
\end{align}
where  $m\!\in\! \mathbb{N}$. Here $\e_m\!=\!\hbar\om_0(m+\frac{1}{2})$ represents the single-particle energy-spectrum of $H_0$ and $a_m$ ($a_m^{\dagger}$) is the bosonic annihilation (creation) operator for the $m^\mathrm{th}$ energy level. Due to the number conservation, \emph{i.e.,} $\sum_m a^\dagger_m a_m\!=\!N$, we are restricted to work with the canonical ensemble description. 
The central repercussion of this constraint is that it induces correlations between the occupation numbers of different single-particle occupation states. However, one can still compute the canonical partition function  $Z_N$ for $N$ bosonic particles  via a set of recursion relations starting from $N\!=\!1$ \cite{Barghathi,MAGNUS2017,Mullin.W.J.,Giraud,Borrmann,canonicalWicks,Satya.N.Majumdar},
\begin{align}
 Z_N=&\frac{1}{N}\sum_{k=1}^N Z_1(k\b)Z_{N-k}\label{partition1}.
 \end{align}
Here, $Z_1(k\b)\!=\!\sum\nolimits_m e^{-k\b\e_m}$ represents the partition function for a single-particle state. For our model example, the Green's function $G^>(\omega)$ depends on the two-point correlation involving the occupation number operator. Let $n_m\!=\!a_m^{\dagger} a_m$ be the occupation number operator corresponding to the $m^\mathrm{th}$ energy level, and $\mbar{\cdot\cdot}$ represents thermal averaging with respect to the canonical $N$-particle density matrix 
\begin{align}
    \rho_N=\frac{1}{Z_N}e^{-\b\sum\nolimits_m \e_m n_m}\delta_{\sum\nolimits_m n_m,N}.
\end{align}
The average occupation of $m^\mathrm{th}$ level can be computed using the following recursion relations \cite{Barghathi}
\begin{align}
    \mbar{n}_m=&\frac{1}{Z_N}\sum_{k=1}^N e^{-\b\e_m k}Z_{N-k},\label{occu-2}\\
     \mmbar{n}_m=&\frac{Z_N}{Z_{N+1}}e^{-\b\e_m}\big(1+\mbar{n}_m\big).\label{occu-1}
\end{align}
 starting from $Z_0\!=\!1$ and $\mzbar{n}_m\!=\!0$. 

In order to calculate the Onsager transport coefficients, the key non-trivial quantity required is the greater Green's function, given by
\begin{align}
    G&^>_N(\om)=-2\pi\frac{\om_0^2}{4} \sum\nolimits_m (m\!+\!1)(m\!+\!2)\Big[\delta(\om\!-\!2\om_0)\big(\mbar{n}_m\nonumber\\
    &\quad\quad+\mbar{n_m n}_{m+2}\big)+\delta(\om\!+\!2\om_0)\big(\mbar{n}_{m+2}+\mbar{n_{m+2} n}_{m}\big)\Big]\nonumber\\
    &-2\pi\frac{1}{\hbar^2}\sum\nolimits_{m,m'}\e_m\e_{m'}\big(\mbar{n_m n}_{m'}-\mbar{n}_m\mbar{n}_{m'}\big)\delta(\om).\label{G_N}
\end{align}
 The challenge of calculating the two-level occupation correlation can be circumvented by using the relation  \cite{Barghathi,MAGNUS2017,Mullin.W.J.,Giraud,Borrmann,canonicalWicks,Satya.N.Majumdar} 
\begin{align}
    \mbar{n_m n}_{m'}=-\frac{e^{\b\e_m}\mbar{n}_m-e^{\b\e_m'}\mbar{n}_{m'}}{e^{\b\e_m}-e^{\b\e_m'}}\label{occu-cor},
\end{align}
which simplifies the expression for the Green's function in Eq.~\eqref{G_N},
\begin{align}
    G&^>_N(\om)=-2\pi\frac{\om_0^2}{4} \sum\nolimits_m (m\!+\!1)(m\!+\!2)\big(\mbar{n}_{m}-\mbar{n}_{m+2}\big)\nonumber\\
    &\quad\quad\big[\big\{1+n_B(2\om_0)\big\}\delta(\om\!-\!2\om_0)+n_B(2\om_0)\delta(\om\!+\!2\om_0)\big]\nonumber\\
    &-2\pi\frac{1}{\hbar^2}\sum\nolimits_{m,m'}\e_m\e_{m'}\big(\mbar{n_m n}_{m'}-\mbar{n}_m\mbar{n}_{m'}\big)\delta(\om)\nonumber\\
    &=-2\pi\frac{\om_0}{\hbar}\sum\nolimits_m\e_m\mbar{n}_m\big[\big\{1+n_B(2\om_0)\big\}\delta(\om\!-\!2\om_0)\nonumber\\
    &\quad\quad\quad\quad\quad\quad\quad +n_B(2\om_0)\delta(\om\!+\!2\om_0)\big]\nonumber\\
    &-2\pi\frac{1}{\hbar^2}\sum\nolimits_{m,m'}\e_m\e_{m'}\big(\mbar{n_m n}_{m'}-\mbar{n}_m\mbar{n}_{m'}\big)\delta(\om),\label{G_Nn}
\end{align}
where $n_B(\om)=[e^{\b\hbar\om}-1]^{-1}$ is the Bose-Einstein distribution function. The two summations in the above expression correspond to the average energy and the variance of energy of the unperturbed Hamiltonian $H_0$.

Finally, we derive the formal expressions for the Onsager transport coefficients as follows,
\begin{align}
    \mathcal{L}_{ww}\!=&\frac{1}{\tau_{cyc}}\Big[\llangle E^2\rrangle+\frac{\la{E}\ra}{\b}F(2\om_0)\Big],\label{modelLww}\\
    \mathcal{L}_{wq}\!=&\mathcal{L}_{qw}\!=\!-\frac{1}{\tau_{cyc}}\Big[\llangle E^2\rrangle\!-\!\l\b\llangle E^3\rrangle\Big],\label{modelLwq}\\
    \mathcal{L}_{qq}\!=&\frac{1}{\tau_{cyc}}\Big[(1+2\l)\llangle E^2\rrangle\!-\!\l\b\Big\{\llangle E^3\rrangle+\frac{\l}{2}\big(\b\llangle E^4\rrangle\nonumber\\&\quad\quad\quad\quad-\!3\llangle E^3\rrangle\big)\Big\}\Big],\label{modelLqq}
\end{align}
where the $n^\mathrm{th}$ cumulant of energy $\llangle E^n\rrangle$ corresponding to $H_0$ is given by 
\begin{align}
    \llangle E^n\rrangle=(\!-\!1)^n\frac{\partial^n}{\partial\b^n}\ln{ Z_N}.
\end{align}
Additionally, the variance of the total work current can be computed following Eq.~\eqref{varjw} and is given by 
\begin{align} 
\frac{\llangle\mathcal{J}_w^2\rrangle}{2}\!=\!\frac{1}{\tau_{cyc}}\Big[\llangle E^2 \rrangle+\frac{\la E\ra}{\b}(\b\h\om_0)\coth{\big(\b\h\om_0\big)}F(2\om_0)\Big].\label{model-wfluc}
\end{align}
Notably, the breakdown of the traditional work FDR in this model is evident from the fact that $\coth (x)>\frac{1}{x}$ for $x>0$,  as reported in Eq.~\eqref{w fdr}. It is also important to mention that the appearance of the higher order cumulant of energy $\llangle E^n\rrangle$ is a consequence of the presence of the interaction term $\lambda H_1$ in the initial total Hamiltonian $H(0)$.
\begin{figure}[H]
    \centering
    \includegraphics[width=1.0\columnwidth]{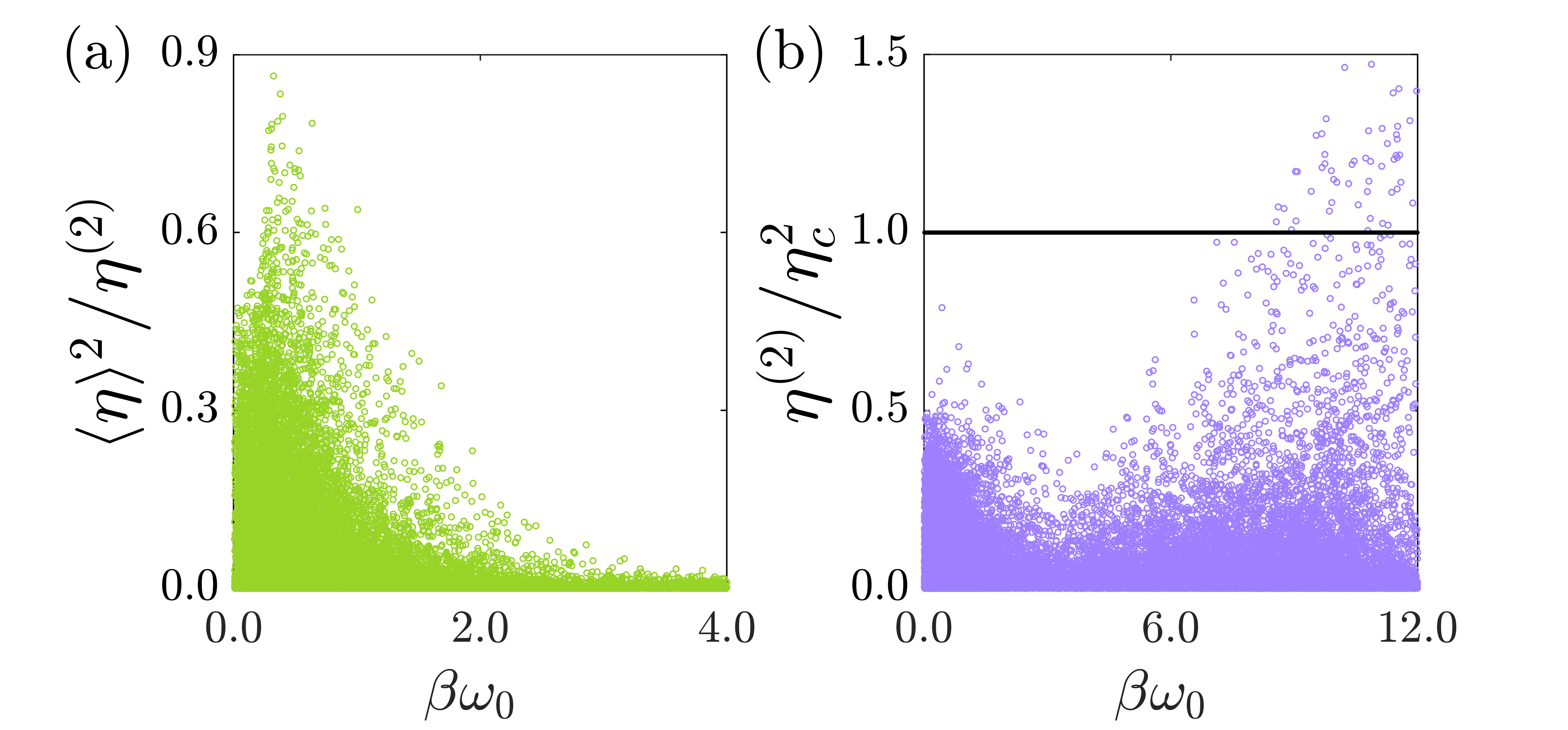}
     \caption{Engine regime: characterized by $\la\mathcal{J}_w\ra<0$, $\la\mathcal{J}_h\ra>0$ (necessary requirement: $\mathcal{F}_q\!>\!\mathcal{F}_w$). Parameters chosen from uniform distributions: $\b\in [0,\, 4]$, $\om_0\in[0,\, 3]$, $\Delta\b  \in[0,\, 0.3\b]$, $\l\in[0,\,0.3]$, and $\Delta\l\in[0,\,0.1]$. Simulations were performed over one million points. (a) Validity of the lower bound {\it i.e.,} $\eta^{(2)} \ge \la \eta\ra^2$ is observed, and (b) Violation of the upper bound  is observed, {\it i.e.,} $\eta^{(2)} \nleq \eta_c^2$.}
     \label{fig1}
\end{figure} 
\begin{figure}[H]
    \centering
\includegraphics[width=1.0\columnwidth]{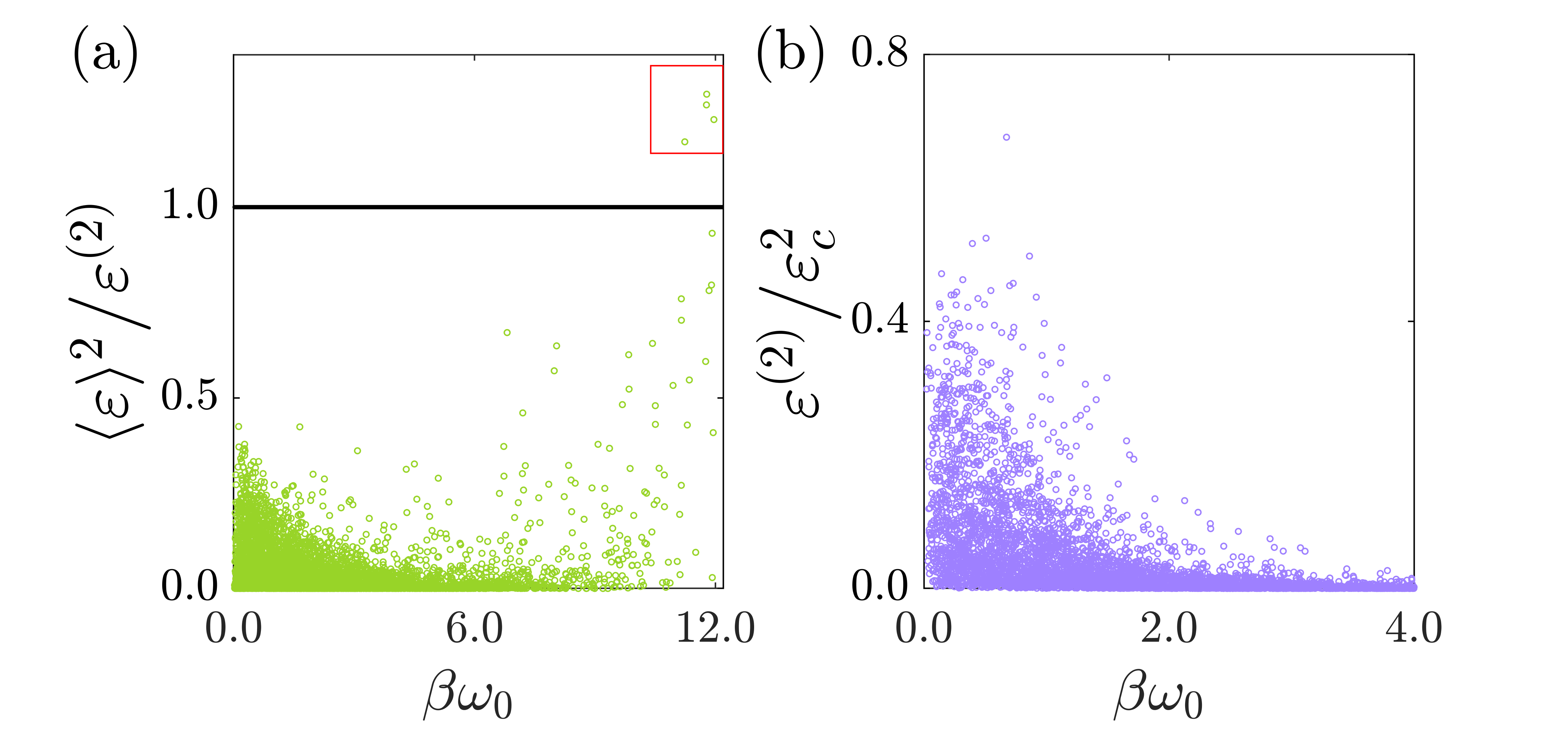}
     \caption{Refrigerator regime: characterized by $\la\mathcal{J}_c\ra>0$, $\la\mathcal{J}_w\ra>0$ (necessary requirement:  $\mathcal{F}_w\!>\!\mathcal{F}_q$). Parameters chosen from uniform distributions: $\b\in [0,\, 4]$, $\om_0\in[0,\, 3]$, $\Delta\b  \in[0,\, 0.3\b]$, $\l\in[0,\,0.3]$, and $\Delta\l\in[0,\,0.1]$. Simulations were performed over one million points. (a) Violation of the lower bound {\it i.e.,} $\varepsilon^{(2)} \ngeq \la \varepsilon\ra^2$ is observed (shown in the red box), and (b) validity of the upper bound  is observed, {\it i.e.,} $\varepsilon^{(2)} \le \varepsilon_c^2$.}
     \label{fig2}
\end{figure}

In the following, we present numerical results for $N\!=\!5$ bosons confined in a harmonic trap. Starting from the single particle partition function $Z_1\!=\csch\big(\b\h\om_0/2\big)/2$, we use the recursion relation given in Eq.~\eqref{partition1} to calculate $Z_5$. The various Onsager coefficients are computed numerically using Eqs.~\eqref{modelLww} to \eqref{modelLqq}. We investigate bounds on the quantities $\eta^{(2)}$ [defined in Eq.~\eqref{def-eng}] in the  engine operational regime and  $\varepsilon^{(2)}$ [defined in Eq.~\eqref{def-ref}] in the refrigerator operational regime. In Fig.~\ref{fig1} (a) we display the ratio $\eta^{(2)}/\langle \eta \rangle^2$, which validates the existence of the lower bound, and in contrast to the autonomous continuous engine, the violation of the upper bound on $\eta^{(2)}$ is observed in Fig.~\ref{fig1} (b). Furthermore, we observe the opposite trend in the refrigerator regime, \emph{i.e.}, the violation of the lower bound, and the existence of the upper bound, as shown in Fig.~\ref{fig2} (a) and (b), respectively. These observations precisely match our analytical predictions, and it is worth emphasizing that the central reason for the violation of these bounds in the linear response regime is due to the breakdown of the traditional FDR for work.

\section{Summary}
\label{sec:summary}
In this work, we utilized the Schwinger-Keldysh nonequilibrium Green's function formalism to derive an analytical expression for the joint CGF for total work and heat from the hot reservoir in the quantum Otto cycle with an arbitrary many-body working medium. The derived CGF is valid up to the second order of the driving protocol $\lambda(t)$ and satisfies the heat engine fluctuation relation. This CGF is applicable to arbitrary perturbative protocols and interaction Hamiltonians. Furthermore, we developed a consistent linear response framework by focusing on the first and second cumulants of total work and heat in the limit of small driving amplitude and small temperature difference. Our analysis revealed the violation of the traditional FDR for the work current as a consequence of quantum non-adiabatic driving during the work strokes. However, for the heat current, the traditional FDR remains valid since the heat exchange stroke in the Otto cycle does not directly experience external driving. These findings have significant implications as they establish different universal bounds on fluctuations for the engine and refrigerator regimes. Specifically, in the engine regime, we observed that the ratio of output to input fluctuations is bounded from below, whereas in the refrigerator regime, the opposite trend is observed--the ratio of fluctuations is bounded from above.  These results stand in stark contrast to the behavior observed for autonomous steady-state engines in earlier studies. Moreover, we found that the bounds on the ratio of fluctuations become similar for the two different families of machines only when the Otto cycle is driven in a quantum-adiabatic manner. This highlights the role of quantum effects in determining the bounds on fluctuations for thermal machines. Our study also established connections to the TURs for work and heat currents in the linear response regime, providing further insight into the thermodynamic behavior of the system. Future work will be directed toward analyzing the bounds beyond the linear response regime. It will be intriguing to extend the above analysis to other classes of quantum thermal machines and explore the implications of quantum effects on the performance and thermodynamic behavior of these systems. 

\section*{ACKNOWLEDGMENTS}
S.M. acknowledges financial support from the CSIR, India (File number: 09/936(0273)/2019-EMR-I). B.K.A. acknowledges the MATRICS grant MTR/2020/000472 from SERB, Government of India. B.K.A. also thanks the Shastri Indo-Canadian Institute for providing financial support for this research work in the form of a Shastri Institutional Collaborative Research Grant (SICRG). 

\onecolumngrid
\renewcommand{\theequation}{A\arabic{equation}}

\renewcommand{\thesection}{A\arabic{section}}
\setcounter{equation}{0}
\appendix
\section{Joint CF on Keldysh contour:}
\label{AppA}
In this Appendix, we present the details of obtaining the joint CF given in Eq.~\eqref{contourCF} in the main text.  Recall that, the CF is the Fourier transform of the joint PD obtained by performing two-time measurements at the end-points of each stroke of the Otto cycle, 
\begin{align}
    \chi(\gamma_w ,\gamma_h)={\mathrm{Tr}\big[{\cal U}_\mathrm{e}^\dagger e^{(i\gamma_w-i\gamma_h)H(\tau)}{\cal U}_\mathrm{e}e^{(-i\gamma_w-\beta_c) H(0)}\big]}\times{\mathrm{Tr}\big[{\cal U}_\mathrm{c}^\dagger e^{i\gamma_w H(0)}{\cal U}_\mathrm{c}e^{(-i\gamma_w+i\gamma_h-\beta_h) H(\tau)}\big]}{{\cal Z}^{-1}_h[H(\tau)]}\,{{\cal Z}^{-1}_c[H(0)]}.\label{CGFapp}
\end{align}
Here, the Hamiltonian of the working medium  at any instant $t \in [0,\tau]$ is given by
 $H(t)\!=\!H_0+\l(t)H_1$, where $\l(0)\! =\! \l_0$ and $\l(\tau)\!=\! \l_1$.
In time-dependent perturbation theory, we express all the evolution operators in the interaction picture with respect to $H_0$. Here, we consider the first trace of Eq.~\eqref{CGFapp} only. A similar analysis applies to the second trace as well. For the sake of simplicity, we are dropping the $\mathrm{e}$ subscript from ${\cal U}_\mathrm{e}$. In the interaction picture. We obtain
\begin{gather}
    {\cal U}_I(t)={\cal T}_+\exp\Big[-\frac{i}{\hbar}\int_0^t \l(s)H_1^I(s)\, ds\Big],\label{1st-u}\\
    {\cal U}^{(0)}_I(t)={\cal T}_+\exp\Big[-\frac{i}{\hbar}\int_0^t \l_0H_1^I(s)\, ds\Big], \label{2nd-u}\\ 
    {\cal U}^{(\tau)}_I(t)={\cal T}_+\exp\Big[-\frac{i}{\hbar}\int_0^t \l_1 H_1^I(s)\, ds\Big] \label{3rd-u},
\end{gather}
where, the superscripts $(0)$ and $(\tau)$ in Eqs.~\eqref{2nd-u} and \eqref{3rd-u} indicate interaction picture evolution operator with fixed $\l_0$ and $\l_1$, respectively. Here ${\cal T}_{+}$ is the time-ordering operator which arranges the operators from left to right with decreasing time.  Note that, $ H_1^I(s)\!=\!{\cal U}_0^\dagger(s)\,H_1 \, {\cal U}_0(s)$, where
 ${\cal U}_0(t)\!=\!\exp{(\!-\!iH_0 t/\hbar)}$ represents the free evolution operator. We express the first trace in Eq.~\eqref{CGFapp}  as,
\begin{align}
    \mathrm{Tr}&\Big[e^{-i\gamma_w H(0)}\,{\cal U}^\dagger(\tau)\, e^{(i\gamma_w-i\gamma_h)H(\tau)}\,{\cal U}(\tau)\, e^{-\beta_c H(0)}\Big]\nonumber \\
    =&\mathrm{Tr}\Big[{{\cal U}^{(0)}}^\dagger(-\hbar\gw)\, {\cal U}^\dagger(\tau)\, {{\cal U}^{(\tau)}}(-\hbar\gw+\hbar\gh)\,{\cal U}(\tau)\, {{\cal U}^{(0)}}(-i\hbar\beta_c)\Big]\nonumber\\
    =&\mathrm{Tr}\Big[{{\cal U}_I^{(0)}}^\dagger(-\hbar\gw)\,{\cal U}_0^\dagger(-\h\gw)\, {\cal U}_I^\dagger(\tau)\,{\cal U}_0^\dagger(\tau)\, {\cal U}_0(-\h\gw+\h\gh)\,{{\cal U}_I^{(\tau)}}(-\h\gw+\h\gh)\,
   {\cal U}_0(\tau)\, {\cal U}_I(\tau)\, {\cal U}_0(-i\h\beta_c)\,{{\cal U}_I^{(0)}}(-i\h\beta_c)\Big]\nonumber\\
   =&\mathrm{Tr}\Big[{\cal U}_0(-i\h\beta_c)\,{\cal U}_0(\h\gh)\,\overbrace{{\cal U}_0^\dagger(\h\gh)\,{\cal U}_I^{(0)}(-i\h\beta_c)\,{\cal U}_0(\h\gh)}\,\underbrace{{\cal U}^\dagger_0(\h\gh)\,{{\cal U}_I^{(0)}}^\dagger(-\h\gw)\,{\cal U}_0(\h\gh)}\,\nonumber\\&\quad\quad\overbrace{{\cal U}_0^\dagger(-\h\gw+\h\gh) \,{\cal U}_I^\dagger(\tau)\,{\cal U}_0(-\h\gw+\h\gh)} \underbrace{{\cal U}_0^\dagger(\tau)\,{{\cal U}_I^{(\tau)}}(-\h\gw+\h\gh)\,
   {\cal U}_0(\tau)}\, \overbrace{{\cal U}_I(\tau) }\Big]\nonumber\\
   =&\Big\la  {\cal T}_+e^{-\frac{i}{\hbar}\int_{\h\gh}^{\h\gh-i\h\beta_c}\l_0 H_1^I(s)ds}\,{\cal T}_-e^{-\frac{i}{\hbar}\int_{-\h\gw+\h\gh}^{\h\gh}\l_0 H_1^I(s)ds}\,{\cal T}_-e^{-\frac{i}{\hbar}\int_{\tau-\h\gw+\h\gh}^{-\h\gw+\h\gh}\l(s+\gw-\gh) H_1^I(s)ds}\nonumber\\&\quad\quad\,{\cal T}_+e^{-\frac{i}{\hbar}\int_{\tau}^{\tau-\h\gw+\h\gh}\l_1 H_1^I(s)ds} \,{\cal T}_+e^{-\frac{i}{\hbar}\int_{0}^{\tau}\l(s) H_1^I(s)ds}\Big\ra_{\wl{\beta}_c}\mathrm{Tr}\Big[e^{(-\beta_c-i\gh)H_0}\Big]\nonumber\\=& \Big\la {\cal T}_{c}\,e^{-\frac{i}{\hbar} \int_{c}\l_{C}(s) H_1^I(s)\,ds} \Big\ra_{\wl{\beta}_c}\,\mathrm{Tr}\Big[e^{(-\beta_c-i\gh)H_0}\Big].
\end{align}
In the last line, we map the real-time evolution onto a modified Keldysh contour $C$ [see Fig.~\ref{figcontour} (a) in the main text] by introducing the contour-time ordered operator ${\cal T}_C$ with contour-time dependent driving parameter $\l_{C}(s)$. Similarly, we can map the partition function ${\cal Z}_c[H(0)]$  appearing in Eq.~\eqref{CGFapp} onto the same contour $C$. For the partition function, we obtain
\begin{align}
    {\cal Z}_c[H(0)]&=\mathrm{Tr}\Big[e^{-\beta_c H(0)}\Big]=\mathrm{Tr}\Big[ \,{\cal U}_0(-i\h\b_c)\,{\cal U}_I^{(0)}(-i\h\b_c)\,\Big]=\mathrm{Tr}\Big[ \,{\cal U}_0(-i\b_c)\,{\cal U}^\dagger_0(\h\gh)\,{\cal U}_I^{(0)}(-i\h\b_c)\,{\cal U}_0(\h\gh)\,\Big]\nonumber\\&= \Big\la {\cal T}_{c}\,e^{-\frac{i}{\h} \int_{\h\gh}^{\h\gh-i\h\b_c}\l_{0} H_1^I(s)\,ds} \Big\ra_{{\beta}_c}\,\mathrm{Tr}\big[e^{-\beta_c H_0}\big],
\end{align}
where the contour-time dependent first term is mapped onto a Keldysh contour which runs only on the vertical line of Fig.~\ref{figcontour} (a). The same prescription also applies for the second trace and the partition function ${\cal Z}_h[H(\tau)]$ in Eq.~\eqref{CGFapp}, resulting in the second modified contour [see Fig.~\ref{figcontour} (b)].

\renewcommand{\theequation}{B\arabic{equation}}
\setcounter{equation}{0}
\section{Details of the calculation of joint CGF}
\label{AppB}
In this Appendix, we will provide details of the derivation of Eq.~\eqref{main-CGF}. We will closely follow the prescription provided in Ref.~\cite{NEGF-work}. Here, we will demonstrate the series expansion for the first contour of the joint heat and work characteristic function: 
 \begin{align}
   \chi_c(\gamma_w,\gamma_h)&\coloneqq \frac{\big\langle {\cal T}_C \,e^{-\frac{i}{\hbar}\int_C\! \lambda_C(s_1)H_1^I(s_1) ds_1}\big\rangle_{\wl{\b}_c}}{\big\langle {\cal T}_C\, e^{-\frac{i}{\hbar}\int_{\hbar\gamma_h}^{\hbar\gamma_h\!-\!i\hbar\beta_c}\! \lambda_0 H_1^I(s_1) ds_1}\big\rangle_{{\b}_c}}\nonumber\\
   &=\frac{1+\displaystyle \sum_{n=1}^\infty\bigg[\prod_{j=1}^n\int_C ds_j \l_C(s_j)  \Theta_C(s_j\!-\!s_{j+1}) \left(\!-\frac{i}{\hbar}\right)^n\big\la H_1^I(s_1)...H_1^I(s_n)\big\ra_{\wl{\b}_c}\bigg]}{1+\displaystyle \sum_{n=1}^\infty\bigg[\prod_{j=1}^n\int_{\hbar\gh}^{\hbar\gh-i\hbar\b_c} ds_j \l_C(s_j)  \Theta_C(s_j\!-\!s_{j+1}) \left(\!-\frac{i}{\hbar}\right)^n\big\la H_1^I(s_1)...H_1^I(s_n)\big\ra_{{\b}_c}\bigg]}, \label{CF1}
\end{align}
where, the contour step function $\Theta_C(s_j\!-\!s_{j+1})$ is appearing because of the contour-time ordering. A more convenient way is to consider the expansion of the cumulant generating function $\ln[\chi_c(\gw,\gh)]$, 
where the series expansion is expressed in terms of $n$-point connected correlation functions  (cumulant correlation functions),
\begin{align}
    \ln\, \chi_c(\gw,\gh)= \sum_{n=1}^\infty\bigg[\prod_{j=1}^n\int_C d\mathbf{S}_j \wl{G}_c(s_1,...,s_n)-\prod_{j=1}^n\int_{\hbar\gh}^{\hbar\gh-i\hbar\b_c} d\mathbf{S}_j {G}_c(s_1,...,s_n)\bigg],\label{cgf1}
\end{align}
where, $d\mathbf{S}_j=ds_j \l_C(s_j)  \Theta_C(s_j\!-\!s_{j+1})$ and
\begin{align}
    \wl{G}_c(s_1,...,s_n)&=\left(\!-\frac{i}{\hbar}\right)^n \!\llangle H_1^I(s_1)...H_1^I(s_n)\rrangle_{\wl{\beta}_c},\nonumber\\{G}_c(s_1,...,s_n)&=\left(\!-\frac{i}{\hbar}\right)^n \!\llangle H_1^I(s_1)...H_1^I(s_n)\rrangle_{{\beta}_c}.
\end{align}
Note that, $\llangle .\rrangle$ represents cumulant correlation function [see Eq.~\eqref{llangle} in the main text]. Crucially, the contour-time dependent driving parameter $\l_C(s)$ is piece-wise defined along the contour $C$:
\begin{align}
  \l(s)=
    \begin{cases}
     \l(s) & \mathrm{Part }\;1:\;s\in [0,\tau),\\
     \l_1 & \mathrm{Part }\;2:\;s\in [\tau,\tau\!-\!\h\gw+\h\gh),\\
     \l(s+\h\gw\!-\!\h\gh) & \mathrm{Part }\;3:\;s\in [\tau\!-\!\h\gw+\h\gh,-\h\gw+\h\gh),\\
     \l_0 & \mathrm{Part }\;4:\;s\in [-\h\gw+\h\gh,\h\gh\!-\!i\h\b_c].
    \end{cases}
    \label{partC}
\end{align}
In what follows, we will calculate the right-hand side of Eq.~\eqref{cgf1} up to the second order of the driving parameter $\l(s)$. The contribution to the first order of {$\l(s)$} is given by
\begin{align}
    i\big[\gw(\l_1\!-\!\l_0)-\gh \l_1\big]\la H_1\ra_{\wl{\b}_c}-\b_c\l_0\big[\la H_1\ra_{\wl{\b}_c}\!-\!\la H_1\ra_{{\b}_c}\big].
\end{align}
Next, for  the contribution to the second order in {$\l(s)$} we have to calculate  double integrals given by
\begin{align}
    &\int_C \!\!ds_1\,\l_C(s_1)\!\int_C\! \!ds_2\,\l_C(s_2)\,\Theta_C(s_1\!-\!s_2)\, \wl{G}_c(s_1,s_2)-\!\!\int_{\h\gh}^{\h\gh\!-\!i\h\b_c}\!\! ds_1\,\l_C(s_1)\!\int_{\h\gh}^{\h\gh\!-\!i\h\b_c}\!\! ds_2\,\l_C(s_2)\,\Theta_C(s_1\!-\!s_2)\,G_c(s_1,s_2)\nonumber\\
    =&\int_C\!\! ds_1\,\l_C(s_1)\!\!\int_C\! ds_2\,\l_C(s_2)\,\Theta_C(s_1\!-\!s_2)\, \wl{G}_c^>(s_1\!-\!s_2)-\!\!\int_{\h\gh}^{\h\gh\!-\!i\h\b_c}\! \!ds_1\,\l_C(s_1)\!\!\int_{\h\gh}^{\h\gh\!-\!i\h\b_c}\!\! ds_2\,\l_C(s_2)\,\Theta_C(s_1\!-\!s_2) \,G_c^>(s_1\!-\!s_2).
    \label{2nd order}
\end{align}
Note that the greater Green's functions are introduced in Eq.~\eqref{g>} in the main text. Due to Eq.~\eqref{partC}, the double integral along the contour $C$ is expressed as a sum of double integrals of parts $(i,j)$, where $i,j\in\{1,2,3,4\}$. Importantly, once we take into account the contour step function $\Theta(s_1\!-\!s_2)$, only 10 out of the possible 16 terms survive in this sum:
\begin{align}
    (i,j)=\begin{cases} 
        (1,1),\\
        (2,1),\;(2,2),\\
        (3,1),\;(3,2),\;(3,3),\\
        (4,1),\;(4,2),\;(4,3),\;(4,4).
    \end{cases}
    \label{surviving parts}
\end{align}
Bellow we show details of calculating a few of these double integrations: 

\noindent1. For $(i,j)= (4,1)$
\begin{align}
    &\int \frac{d\om}{2\pi}\int_{-\!\h\gw+\h\gh}^{\h\gh\!-\!i\h\b_c}\!ds_1\,\l_0\,e^{-i\om s_1}\int_{0}^{\tau}\!ds_2\,\l(s_2)\,e^{i\om s_2}\wl{G}_c^>(\om)\nonumber\\
    &=\int \frac{d\om}{2\pi}\l_0 \frac{e^{-\h\om(\b_c+i\gh)}\wl{G}_c^>(\om)}{-i\om}\int_{0}^{\tau}\!ds_2\,\l(s_2)\,e^{i\om s_2}-\int \frac{d\om}{2\pi}\l_0 \frac{\wl{G}_c^>(\om)}{-i\om}\,e^{i\h\om(\gw-\gh)}\int_{0}^{\tau}\!ds_2\,\l(s_2)\,e^{i\om s_2}\nonumber\\
    &=\int \frac{d\om}{2\pi}\l_0 \frac{\wl{G}_c^>(-\om)}{-i\om}\int_{0}^{\tau}\!ds_2\,\l(s_2)\,e^{i\om s_2}+\int \frac{d\om}{2\pi}\l_0 \frac{\wl{G}_c^>(\om)}{i\om}\,e^{i\h\om(\gw-\gh)}\int_{0}^{\tau}\!ds_2\,\l(s_2)\,e^{i\om s_2}\nonumber\\
    &=\int \frac{d\om}{2\pi} \frac{\wl{G}_c^>(\om)}{i\om}\,\l_0\int_{0}^{\tau}\!ds_2\,\l(s_2)\,e^{-i\om s_2}+\int \frac{d\om}{2\pi} \frac{\wl{G}_c^>(\om)}{i\om}\,\l_0\,e^{i\h\om(\gw-\gh)}\int_{0}^{\tau}\!ds_2\,\l(s_2)\,e^{i\om s_2}.
\end{align}
where to go from the second step to the third step we have used the KMS relation  given in Eq.~\eqref{kms1} in   the main text.

\noindent 2. For $(i,j)= (4,2)$
\begin{align}
    &\int \frac{d\om}{2\pi}\int_{-\!\h\gw+\h\gh}^{\h\gh\!-\!i\h\b_c}\!ds_1\,\l_0\,e^{-i\om s_1}\int_{\tau}^{\tau\!-\!\h\gw+\h\gh}\!ds_2\,\l_1\,e^{i\om s_2}\wl{G}_c^>(\om)\nonumber\\
    &=-\int \frac{d\om}{2\pi}\l_0\l_1 \frac{e^{-\h\om(\b_c+i\gh)}\wl{G}_c^>(\om)}{\om^2}\big(1-e^{-i\h\om(\gw-\gh)}\big)\,e^{i\om\tau}-\int \frac{d\om}{2\pi}\l_0\l_1 \frac{\wl{G}_c^>(\om)}{\om^2}\big(1-e^{i\h\om(\gw-\gh)}\big)\,e^{i\om\tau}\nonumber\\
    &=-\int \frac{d\om}{2\pi}\l_0\l_1 \frac{\wl{G}_c^>(-\om)}{\om^2}\big(1-e^{-i\h\om(\gw-\gh)}\big)\,e^{i\om\tau}-\int \frac{d\om}{2\pi}\l_0\l_1 \frac{\wl{G}_c^>(\om)}{\om^2}\big(1-e^{i\h\om(\gw-\gh)}\big)\,e^{i\om\tau}\nonumber\\
    &=\int \frac{d\om}{2\pi}\frac{\wl{G}_c^>(\om)}{\om^2} \big[-{2\l_0\l_1}\big(1-e^{i\h\om(\gw-\gh)}\big)\,\cos{(\om\tau)}\big],
\end{align}

\noindent 3. For $(i,j)= (4,3)$
\begin{align}
     &\int \frac{d\om}{2\pi}\int_{-\!\h\gw+\h\gh}^{\h\gh\!-\!i\h\b_c}\!ds_1\,\l_0\,e^{-i\om s_1}\int_{\tau-\h\gw+\h\gh}^{-\h\gw+\h\gh}\!ds_2\,\l(s_2+\h\gw-\h\gh)\,e^{i\om s_2}\wl{G}_c^>(\om)\nonumber\\
     &=-\int \frac{d\om}{2\pi}\int_{-\!\h\gw+\h\gh}^{\h\gh\!-\!i\h\b_c}\!ds_1\,\l_0\,e^{-i\om s_1}\int_{0}^{\tau}\!ds_2\,\l(s_2)\,e^{i\om (s_2-\h\gw+\h\gh)}\wl{G}_c^>(\om)\nonumber\\
     &=-\int \frac{d\om}{2\pi} \frac{\wl{G}_c^>(\om)}{i\om}\,\l_0\,e^{i\h\om(\gw-\gh)}\int_{0}^{\tau}\!ds_2\,\l(s_2)\,e^{-i\om s_2}-\int \frac{d\om}{2\pi} \frac{\wl{G}_c^>(\om)}{i\om}\,\l_0\int_{0}^{\tau}\!ds_2\,\l(s_2)\,e^{i\om s_2}.
\end{align}
The intermediate steps are similar to the  (4,1) case.

\noindent 4. For $(i,j)= (4,4)$
\begin{align}
     &\int \frac{d\om}{2\pi}\int_{-\!\h\gw+\h\gh}^{\h\gh\!-\!i\h\b_c}\!ds_1\,\l_0\,e^{-i\om s_1}\int_{-\h\gw+\h\gh}^{s_1}\!ds_2\,\l_0\,e^{i\om s_2}\wl{G}_c^>(\om)-\int \frac{d\om}{2\pi}\int_{\h\gh}^{\h\gh\!-\!i\h\b_c}\!ds_1\,\l_0\,e^{-i\om s_1}\int_{\h\gh}^{s_1}\!ds_2\,\l_0\,e^{i\om s_2}{G}_c^>(\om)\nonumber\\
     &=\int \frac{d\om}{2\pi}\frac{\wl{G}_c^>(\om)}{\om^2}\l_0^2\big[-i\h\om\gw-\b_c\h\om+1-e^{-\b_c\h\om}e^{-i\h\om\gw}\big]-\int \frac{d\om}{2\pi}\frac{{G}_c^>(\om)}{\om^2}\l_0^2\big[-\b_c\h\om+1-e^{-\b_c\om}\big]\nonumber\\
     &=\int \frac{d\om}{2\pi}\l_0^2\Big[\frac{\wl{G}_c^>(\om)}{\om^2}\big(-i\h\om\gw-\b_c\h\om+1\big)-e^{-i\h\om(\gw-\gh)}\frac{\wl{G}_c^>(-\om)}{\om^2}\Big]-\int \frac{d\om}{2\pi}\l_0^2\Big[\frac{{G}_c^>(\om)}{\om^2}\big(-\b_c\h\om+1)-\frac{{G}_c^>(-\om)}{\om^2}\Big]\nonumber\\
     &=\int \frac{d\om}{2\pi}\frac{\wl{G}_c^>(\om)}{\om^2}\l_0^2\big[-i\h\om\gw-\b_c\h\om+1-e^{i\h\om(\gw-\gh)}\big]+\int \frac{d\om}{2\pi}\frac{{G}_c^>(\om)}{\om}\l_0^2\b_c\h.
\end{align}
Note that, the second term here is stemming from the  canonical partition function ${\cal Z}_c[H(0)]$ [second term in Eq.~\eqref{2nd order}]. The rest of the terms in Eq.~\eqref{surviving parts} can be calculated analogously. Now, following Ref.~\cite{NEGF-work}, we will categorize the 10 non-zero terms given in Eq.~\eqref{surviving parts} in 6 groups:
\begin{enumerate}
    \item $\{(2,2)\}$:
    \begin{align}
        \int\frac{d\om}{2\pi}\wl{G}_c^>(\om)\Big[\frac{\l_1^2}{\om^2}\big(i\h\om(\gw-\gh)+1-e^{i\h\om(\gw-\gh)}\big)\Big].\label{(2,2)}
    \end{align}
    \item $\{(1,1)+(3,1)+(3,3)\}$
    \begin{align}
        \int\frac{d\om}{2\pi}\wl{G}_c^>(\om)\,\Big(1-e^{i\h\om(\gw-\gh)}\Big)\int_0^\tau ds_1\int_0^\tau ds_2\, \l(s_1)\l(s_2)e^{i\om(s_1-s_2)}.\label{(1,1)}
    \end{align}
    \item $\{(2,1)+(3,2)\}$
    \begin{align}
        \int\frac{d\om}{2\pi}\wl{G}_c^>(\om)\Big[-\frac{2\l_1}{\om}\Big(1-e^{i\h\om(\gw-\gh)}\Big)\int_0^\tau ds \,\l(s)\sin{[\om(\tau-s)]}\Big].\label{(3,2)}
    \end{align}
    \item $\{(4,3)+(4,1)\}$
    \begin{align}
        \int\frac{d\om}{2\pi}\wl{G}_c^>(\om)\Big[-\frac{2\l_0}{\om}\Big(1-e^{i\h\om(\gw-\gh)}\Big)\int_0^\tau ds \,\l(s)\sin{(\om s)}\Big].\label{(4,1)}
    \end{align}
    \item $\{(4,2)\}$
    \begin{align}
        \int \frac{d\om}{2\pi}{\wl{G}_c^>(\om)} \Big[-\frac{2\l_0\l_1}{\om^2}\Big(1-e^{i\h\om(\gw-\gh)}\Big)\,\cos{(\om\tau)}\Big].\label{(4,2)}
    \end{align}
    \item $\{(4,4)\}$
    \begin{align}
       \int \frac{d\om}{2\pi}{\wl{G}_c^>(\om)}\Big[\frac{\l_0^2}{\om^2}\Big(-i\h\om\gw-\b_c\h\om+1-e^{i\h\om(\gw-\gh)}\Big)\Big]+\int \frac{d\om}{2\pi}{{G}_c^>(\om)}\frac{\h\b_c \l_0^2}{\om}.\label{(4,4)}
    \end{align}
\end{enumerate}
We further collect all the terms with $\big(1-e^{i\h\om(\gw-\gh)}\big)$ 
\begin{align}
     &\int \frac{d\om}{2\pi}{\wl{G}_c^>(\om)}\frac{1-e^{i\h\om(\gw-\gh)}}{\om^2}\Big[\l_1^2+\l_0^2-2\l_0\l_1\cos{(\om\tau)}+\om^2\int_0^\tau ds_1\int_0^\tau ds_2\, \l(s_1)\l(s_2)e^{i\om(s_1-s_2)}\nonumber\\&\quad\quad\quad-2\om\l_1\int_0^\tau ds \,\l(s)\sin{[\om(\tau-s)]}-2\om\l_0\int_0^\tau ds \,\l(s)\sin{(\om s)}\Big]\nonumber\\
     &=\int \frac{d\om}{2\pi}{\wl{G}_c^>(\om)}\frac{1-e^{i\h\om(\gw-\gh)}}{\om^2}\bigg|\l_1e^{i\om\tau}-\l_0-i\om\int_0^\tau ds\,\l(s)\,e^{i\om s}\bigg|^2\nonumber\\
     &=\int \frac{d\om}{2\pi}{\wl{G}_c^>(\om)}\frac{1-e^{i\h\om(\gw-\gh)}}{\om^2}\bigg|\int_0^\tau ds\,\dot{\l}(s)\,e^{i\om s}\bigg|^2\nonumber\\
     &=\int \frac{d\om}{2\pi}{\wl{G}_c^>(\om)}\frac{1-e^{i\h\om(\gw-\gh)}}{\om^2}\,A(\om),
\end{align}
where
    $A(\om)\!=\!\big|\int_0^\tau ds\,\dot{\l}(s)\,e^{i\om s}\big|^2$ encapsulates information on the speed of the driving protocol $\l(s)$.
Finally, summing all the terms given in Eqs.~\eqref{(2,2)} to \eqref{(4,4)}, we obtain
\begin{align}
    \int \frac{d\om}{2\pi}{\wl{G}_c^>(\om)}\bigg[\frac{1}{\om}\Big(i\h\gw(\l_1^2-\l_0^2)-i\h\gh\l_1^2-\h\b_c\l_0^2\Big)+\frac{1-e^{i\h\om(\gw-\gh)}}{\om^2}\,A(\om)\bigg]+\int \frac{d\om}{2\pi}{{G}_c^>(\om)}\frac{\h\b_c \l_0^2}{\om}.
\end{align}
Therefore, the right-hand side of Eq.~\eqref{cgf1} up to second order of driving parameter $\l(s)$ is given by 
\begin{align}
    i\Big(\gw(\l_1\!-\!\l_0)-\gh \l_1\Big)&\la H_1\ra_{\wl{\b}_c}-\b_c\l_0\Big(\la H_1\ra_{\wl{\b}_c}\!-\!\la H_1\ra_{{\b}_c}\Big)+\Big(i\h\gw(\l_1^2-\l_0^2)-i\h\gh\l_1^2-\h\b_c\l_0^2\Big)\int \frac{d\om}{2\pi}\frac{{\wl{G}_c^>(\om)}}{\om}\nonumber\\&+\h\b_c \l_0^2\int \frac{d\om}{2\pi}\frac{{{G}_c^>(\om)}}{\om}+\int \frac{d\om}{2\pi}{{\wl{G}_c^>(\om)}}\frac{1-e^{i\h\om(\gw-\gh)}}{\om^2}\,A(\om).
\end{align}
It is essential to highlight that the contour $C$ corresponding to the first trace in Eq.~\eqref{CF} in the main text,  contains all the contributions to the statistics of total work and input heat that are solely connected to the cold reservoir inverse temperature $\b_c$. An analogous calculation can be done for the second contour $C'$, which captures all the contributions that are connected to the hot reservoir inverse temperature $\b_h$. 

\renewcommand{\theequation}{C\arabic{equation}}
\setcounter{equation}{0}
\section{Proof for heat FDR}
\label{AppC}
In this Appendix, we will demonstrate the proof for the heat FDR given by Eq.~\eqref{heat-FDR} in the main text. We start with Eq.~\eqref{q-fluc} and take the limit of $\mathcal{F}_w\!=\!0$ and $\mathcal{F}_q\!=\!0$. This immediately hands over the following expression for heat-fluctuation 
\begin{align}
    \llangle \mathcal{J}_h^2 \rrangle\Big|_{\substack{\mathcal{F}_w=0\\ \mathcal{F}_q=0}} \!&= \frac{1}{\tau_{cyc}}\bigg[ \llangle H_0^2\rrangle_{\b_c}\!+\llangle H_0^2\rrangle_{\b_h}\!+2\l_1\big[\llangle H_0 H_1\rrangle_{\b_c}\!\!+\llangle H_0 H_1\rrangle_{\b_h}\big]+2\h\l_1^2\!\!\int\!\frac{d\om}{\pi}\frac{G_c^{'}(\om)+G_h^{'}(\om)}{\om}\!-\!\b_c\l_0\Big[\llangle H_0^2 H_1\rrangle_{\b_c}\nonumber\\
   & \!\!+\h\l_0\!\!\int\!\frac{d\om}{2\pi}\frac{G_c^{''}(\om)}{\om}\Big]\!-\!\b_h\l_1\!\Big[\llangle H_0^2 H_1\rrangle_{\b_h}\!\!+\h\l_1\!\!\int\!\frac{d\om}{2\pi}\frac{G_h^{''}(\om)}{\om}\Big]\!+\!2\h\!\!\int\!\frac{d\om}{2\pi}\frac{G^{'}_c(\om)A(\om)}{\om}+\h^2\!\!\int\!\frac{d\om}{2\pi}{A(\om) G_c^>(\om)}\bigg]_{\substack{\mathcal{F}_w=0\\ \mathcal{F}_q=0}}\nonumber\\
   &=\frac{2}{\tau_{cyc}}\Big\{-\frac{\partial}{\partial\b}\Big[\la H_0\ra_\b+2\l\la H_1\ra_\b +2\h\l^2\!\!\int\frac{d\om}{2\pi}\frac{G^>(\om)}{\om}\Big]-\b\l\frac{\partial^2}{\partial\b^2}\Big[\la H_1\ra_\b+\h\l\!\!\int\frac{d\om}{2\pi}\frac{G^>(\om)}{\om}\Big]\Big\}\nonumber\\
   &=2\cdot\mathcal{L}_{qq},
\end{align}
where we have used $\l\!=\!\l_0$ and $\b\!=\!\b_c$, as adopted in the main text.

\renewcommand{\theequation}{D\arabic{equation}}
\setcounter{equation}{0}
\section{Connection with the TURs}
\label{AppD}
\begin{figure}[H]
    \centering
     \includegraphics[width=0.5\columnwidth]{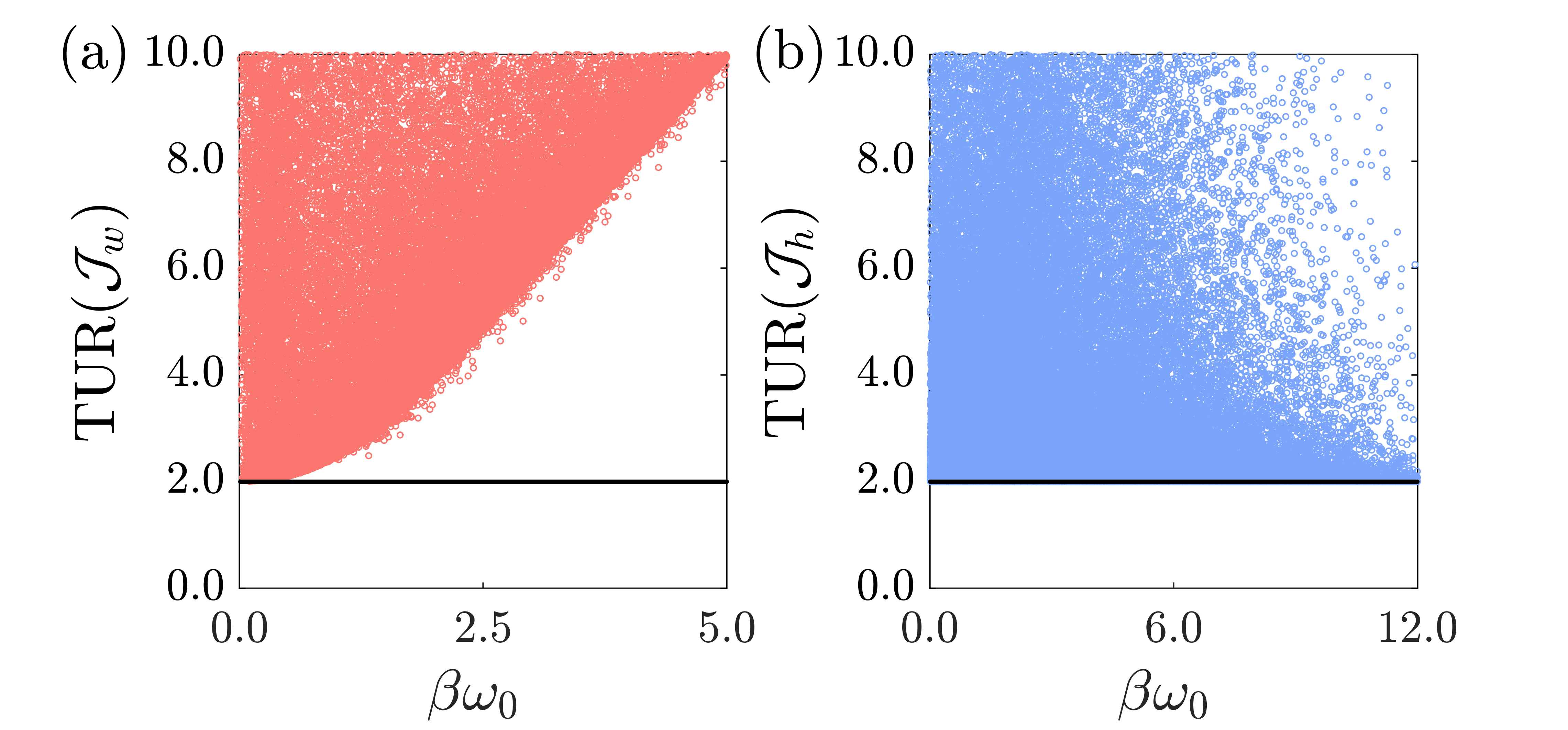}
     \caption{Thermodynamic Uncertainty Relations:  $\mathrm{TUR}(\mathcal{J}_i)\!=\!\la \sigma\ra \llangle \mathcal{J}_i^2\rrangle/\la\mathcal{J}_i\ra^2$, where (a) $i\!=\!w$, and (b) $i\!=\!h$. These results correspond to the example presented in the main text, \emph{i.e.}, N=5 bosons trapped in a harmonic potential. Parameters chosen from uniform distributions: $\b\in [0,\, 4]$, $\om_0\in[0,\, 3]$, $\Delta\b  \in[0,\, 0.3\b]$, $\l\in[0,\,0.3]$, and $\Delta\l\in[0,\,0.1]$. Simulations were done over one million points.}
     \label{fig3}
\end{figure}

In this Appendix, we will provide proof for the thermodynamic uncertainty relations,
\begin{align}
\langle \sigma \rangle \frac{\llangle\mathcal{J}_w^2\rrangle}{\langle \mathcal{J}_w \rangle^2} \geq 2,\quad  
\langle \sigma \rangle \frac{\llangle\mathcal{J}_h^2\rrangle}{\langle \mathcal{J}_h \rangle^2} \geq 2,
\end{align}
where the entropy production rate $\sigma\!=\!{\cal F}_w\la \mathcal{J}_w\ra+{\cal F}_q\la \mathcal{J}_h\ra$. Using Eqs.~\eqref{onsagar} and \eqref{w fdr} we show that
\begin{align}
    \langle \sigma \rangle \frac{\llangle\mathcal{J}_w^2\rrangle}{\langle \mathcal{J}_w \rangle^2}&\ge\la \sigma\ra \frac{2\cdot{\cal L}_{ww}}{\la \mathcal{J}_w\ra^2}\nonumber\\
    &=\big({\cal F}_w\la \mathcal{J}_w\ra+{\cal F}_q\la \mathcal{J}_h\ra\big)\frac{2{\cal L}_{ww}}{\la \mathcal{J}_w\ra^2}\nonumber\\&=2\Big({\cal L}_{ww}^2{\cal F}_{w}^2+2{\cal L}_{ww}{\cal L}_{wq}{\cal F}_{w}{\cal F}_{q}+{\cal L}_{ww}{\cal L}_{qq}{\cal F}_{q}^2\Big)\frac{1}{\la \mathcal{J}_w\ra^2}\nonumber\\
    &=2\Big(\la\mathcal{J}_w\ra^2+\big[{\cal L}_{ww}{\cal L}_{qq}-{\cal L}_{wq}^2\big]\Big)\frac{1}{\la \mathcal{J}_w\ra^2}\nonumber\\
    &=2+\frac{2\cdot\mathrm{det}| {\cal L}|}{\la \mathcal{J}_w\ra^2}\ge2.
\end{align}
The proof of the TUR corresponding to $\mathcal{J}_h$ follows analogously. {In Fig.~\ref{fig3} we display the TURs for work and heat currents for the trapped boson model considered in the main text.}

\renewcommand{\theequation}{E\arabic{equation}}
\setcounter{equation}{0}
\section{Proof for  \mathinhead{\eta_c^2\geq\frac{{\cal F}_w^2\mathcal{ L}_{w w}}{\beta^2{\cal L}_{q q}}\geq\langle \eta \rangle^2}{eng-inequality}}
\label{AppE}
In this Appendix, we will prove the  inequalities stated in Eq.~\eqref{eng-ons} in the main text step by step,
\begin{align}
    \eta_c^{2}\ge\frac{{\cal F}_w^2{\cal L}_{w w}}{\b^2{\cal L}_{q q}}\ge \la \eta\ra^2.\label{eng-bound1}
\end{align}
Let us first consider the first inequality.  We will consider a modified version: $\eta_c^2\b^2\mathcal{L}_{qq}\!-\!\mathcal{L}_{ww}\mathcal{F}_w^2\ge0$. By using the fact that $\eta_c\!=\!\Delta\b/\b\!=\!\mathcal{F}_q/\b$, we observe
\begin{align}
    \eta_c^2\b^2\mathcal{L}_{qq}\!-\!\mathcal{L}_{ww}\mathcal{F}_w^2=&\mathcal{L}_{qq}\mathcal{F}_q^2\!-\!\mathcal{L}_{ww}\mathcal{F}_w^2\nonumber\\
    =&\mathcal{L}_{qq}\mathcal{F}_q^2+\mathcal{L}_{wq}\mathcal{F}_w\mathcal{F}_q\!-\!\mathcal{L}_{wq}\mathcal{F}_w\mathcal{F}_q\!-\!\mathcal{L}_{ww}\mathcal{F}_w^2\nonumber\\
    =&\mathcal{F}_q\la \mathcal{J}_h\ra\!-\!\mathcal{F}_w\la \mathcal{J}_w\ra\nonumber\\
    =&\mathcal{F}_q\la \mathcal{J}_h\ra\Big[1+\frac{\la \eta\ra}{\eta_c}\Big]\nonumber\\
    \ge&0.
\end{align}
In the above steps, we used the engine condition: $\la\mathcal{J}_w\ra\!<\!0$, and $\la \mathcal{J}_h\ra\!>\!0$. Next, let's move on to the second inequality. Our strategy will be to consider a modified version of the second inequality: $\mathcal{L}_{ww}\la \mathcal{J}_w\ra^2\!-\!\mathcal{L}_{qq}\la \mathcal{J}_h\ra^2\!\ge\!0$. We observe 
\begin{align}
    \mathcal{L}_{ww}\la \mathcal{J}_w\ra^2\!-\!\mathcal{L}_{qq}\la \mathcal{J}_h\ra^2&=\big(\mathcal{L}_{ww}\mathcal{L}_{qq}\!-\!\mathcal{L}_{wq}^2\big)\Big[\mathcal{L}_{qq}\mathcal{F}_q^2\!-\!\mathcal{L}_{ww}\mathcal{F}_w^2\Big]\nonumber\\&=\mathrm{det}|\mathcal{L}|\Big[\mathcal{F}_q\la \mathcal{J}_h\ra\!-\!\mathcal{F}_w\la \mathcal{J}_w\ra\Big]\nonumber\\&=\mathrm{det}|\mathcal{L}|\,\mathcal{F}_q\la \mathcal{J}_h\ra\Big[1+\frac{\la \eta\ra}{\eta_c}\Big]\nonumber\\
    &\ge0.
\end{align}
It is important to note that, the two ingredients required for this proof are-
\begin{enumerate}
    \item The positive semi-definiteness of the Onsager matrix, which implies that $\mathrm{det} |\mathcal{L}|\!\ge\!0$, and
    \item The characterization of engine operational regime  by $\la \mathcal{J}_w\ra\!<\!0$ and $\la \mathcal{J}_h\ra\!>\!0$.
\end{enumerate}

\renewcommand{\theequation}{F\arabic{equation}}
\setcounter{equation}{0}
\section{Calculation of CGF for heat from the cold reservoir}
\label{AppF}
 The CF for $q_c$ is obtained by setting $\gh\!=\!\gw\!=\!-\gamma_c$ in Eq.~\eqref{CF}. The new counting variable $\gamma_c$ corresponds to the stochastic variable $(-w-q_h)$,
 \begin{align}
     \chi(\gamma_c)=\int\!\!\!\int d{w}\,d{q_h}\, P({w},{q_h})\,e^{i\gamma_c{(-w-q_h)}}.\label{CF_qc}
 \end{align}
Interestingly, under the perfect thermalization condition with respect to heat reservoirs, the first law of thermodynamics holds even in the stochastic level, \emph{i.e.}, $q_c=-w-q_h$. This is rigorously proven in Appendix A of our previous work \cite{sandipan02}. As a result, the CF in Eq.~\eqref{CF_qc} correctly provides the statistics of the exchanged heat $q_c$ with respect to the cold reservoir. Finally, setting $\gh\!=\!\gw\!=\!-\gamma_c$ in Eq.~\eqref{main-CGF}, we obtain the CGF for $q_c$,
\begin{align}
    \ln &\chi(\gamma_c)=-i\gamma_c\lambda_0\big[\langle H_1\rangle_{\wl{\beta}_h}
     \!\!-\!\langle H_1\rangle_{\wl{\beta}_c}\big]\! -\!i\gamma_c\h\lambda_0^2\!\!\int\!\frac{d\omega}{2\pi}\frac{\wl{G}_h^>(\omega)\!-\!\wl{G}_c^>(\omega)}{\omega}\!-\!\beta_c\lambda_0\Big[\langle H_1\rangle_{\wl{\beta}_c}\!\!-\!\langle H_1\rangle_{\beta_c}+\h\lambda_0\!\!\int\!\frac{d\omega}{2\pi}\frac{\wl{G}_c^>(\omega)\!-\!{G}_c^>(\omega)}{\omega}\Big]\nonumber\\
     &
     \!-\!\beta_h\lambda_1\Big[\langle H_1\rangle_{\wl{\beta}_h}\!\!-\!\langle H_1\rangle_{\beta_h}+\h\lambda_1\!\!\int\!\frac{d\omega}{2\pi}\frac{\wl{G}_h^>(\omega)\!-\!{G}_h^>(\omega)}{\omega}\Big]+\int\!\frac{d\omega}{2\pi}\frac{1\!-\!e^{-i\h\omega\gamma_c}}{\omega^2}{A}(\omega)\wl{G}_h^>(\omega)+\ln \langle e^{i\gamma_c H_0}\rangle_{\beta_c}+\ln \langle e^{-i\gamma_c H_0}\rangle_{\beta_h} .
     \label{CGF_qc}
\end{align}
A crucial point to note here is that now the shifted inverse temperatures  become $\wl{\b}_c\!=\!\b_c\!-\!i\gamma_c$ and $\wl{\b}_h\!=\!{\b}_h+i\gamma_c$, and consequently, $\wl{G}_h^>(w)$ and $\wl{G}_c^>(w)$ are redefined accordingly.

\renewcommand{\theequation}{G\arabic{equation}}
\setcounter{equation}{0}
\section{Proof for \mathinhead{\ve_c^{2} \geq \frac{\b^2{\cal L}_{q q}}{{\cal F}_w^2{\cal L}_{w w}} \geq \la \ve\ra^2}{ref-inequality}}
\label{AppG}
In this Appendix, we will demonstrate the proof for the inequalities in Eq.~\eqref{ref-ons} in the main text,  
\begin{align}
    \ve_c^{2}\ge\frac{\b^2{\cal L}_{q q}}{{\cal F}_w^2{\cal L}_{w w}}\ge \la \ve\ra^2.\label{lower-bound1}
\end{align}
To prove the first inequality, we start with the modified version of the first inequality  $\mathcal{L}_{ww}\mathcal{F}_w^2\!-\!\b^2\mathcal{L}_{qq}/\ve_c^2\ge0$: 
\begin{align}
    \mathcal{L}_{ww}\mathcal{F}_w^2\!-\!\b^2\mathcal{L}_{qq}/\ve_c^2=&\mathcal{L}_{ww}\mathcal{F}_w^2\!-\!\mathcal{L}_{qq}\mathcal{F}_q^2\nonumber\\
    =&\mathcal{F}_w\la \mathcal{J}_w\ra\!-\!\mathcal{F}_q\la \mathcal{J}_h\ra\nonumber\\
    \approx&\mathcal{F}_w\la \mathcal{J}_w\ra+\mathcal{F}_q\la \mathcal{J}_c\ra\nonumber\\
    =&\mathcal{F}_w\la \mathcal{J}_w\ra\Big[1+\frac{\la \ve\ra}{\ve_c}\Big]\nonumber\\
    \ge&0.
\end{align}
Here, we have used the refrigerator condition: $\la\mathcal{J}_c\ra\!>\!0$, and $\la \mathcal{J}_w\ra\!>\!0$. To prove the second inequality, we start with the modified version of the second inequality as follows: 
\begin{align}
    \mathcal{L}_{qq}\la \mathcal{J}_h\ra^2\!-\!\mathcal{L}_{ww}\la \mathcal{J}_w\ra^2&=\big(\mathcal{L}_{ww}\mathcal{L}_{qq}\!-\!\mathcal{L}_{wq}^2\big)\Big[\mathcal{L}_{ww}\mathcal{F}_w^2\!-\!\mathcal{L}_{qq}\mathcal{F}_q^2\Big]\nonumber\\&=\mathrm{det}|\mathcal{L}|\Big[\mathcal{F}_w\la \mathcal{J}_w\ra+\mathcal{F}_q\la \mathcal{J}_c\ra\Big]\nonumber\\&=\mathrm{det}|\mathcal{L}|\,\mathcal{F}_w\la \mathcal{J}_w\ra\Big[1+\frac{\la \ve\ra}{\ve_c}\Big]\nonumber\\
    &\ge0.
\end{align}
In this case, the two ingredients required for this proof are-
\begin{enumerate}
    \item The positive semi-definiteness of the Onsager matrix, which implies that $\mathrm{det} |\mathcal{L}|\!\ge\!0$, and
    \item The characterization of refrigerator operational regime  by $\la \mathcal{J}_c\ra\!>\!0$ and $\la \mathcal{J}_w\ra\!>\!0$.
\end{enumerate}

\twocolumngrid
\bibliographystyle{apsrev4-1}
\bibliography{Reference.bib}

\end{document}